mvpo
\documentclass[%
 aip,
pop,%
 amsmath,amssymb,
 reprint,%
]{revtex4-1}

\usepackage{graphicx}
\usepackage{dcolumn}
\usepackage{bm}
\DeclareMathOperator\arctanh{arctanh}

\begin{document}

\preprint{AIP/123-QED}

\title[Influence of 3D plasmoid dynamics on the transition from collisional to kinetic reconnection]{Influence of 3D plasmoid dynamics on the transition from collisional to kinetic reconnection}

\author{A. Stanier}
\email{stanier@lanl.gov}
\author{W. Daughton}
\author{A. Le}
\author{X. Li}
\author{R. Bird}

\affiliation{Los Alamos National Laboratory, Los Alamos, New Mexico 87545, USA}

\date{\today}

\begin{abstract}
Within the resistive magnetohydrodynamic model, high-Lundquist number reconnection layers are unstable to the plasmoid instability, leading to a turbulent evolution where the reconnection rate can be independent of the underlying resistivity. However, the physical relevance of these results remains questionable for many applications. First, the reconnection electric field is often well above the runaway limit, implying that collisional resistivity is invalid. Furthermore, both theory and simulations suggest that plasmoid formation may rapidly induce a transition to kinetic scales, due to the formation of thin current sheets. Here, this problem is studied for the first time using a first-principles kinetic simulation with a Fokker-Planck collision operator in 3D. The low-$\beta$ reconnecting current layer thins rapidly due to Joule heating before onset of the oblique plasmoid instability. Linear growth rates for standard ($k_y = 0$) tearing modes agree with semi-collisional boundary layer theory, but the angular spectrum of oblique ($|k_y|>0$) modes is significantly narrower than predicted. In the non-linear regime, flux-ropes formed by the instability undergo complex interactions as they are advected and rotated by the reconnection outflow jets, leading to a turbulent state with stochastic magnetic field. In a manner similar to previous 2D results, super-Dreicer fields induce a transition to kinetic reconnection in thin current layers that form between flux-ropes. These results may be testable within new laboratory experiments.

\end{abstract}

\maketitle

\section{\label{sec:intro}Introduction}

Magnetic reconnection is the change in topology of magnetic field-lines in a highly-conducting plasma. The reconnection associated release of stored magnetic energy into plasma kinetic energy is thought to be important in solar flares~\cite{priest02,su13}, planetary magnetospheres~\cite{dungey61,burch16}, and other astrophysical phenomena. In the laboratory, reconnection is usually associated with sawteeth that can lead to the fast collapse of core pressure profiles~\cite{vongoeler74,kadomtsev75,chapman11}, but it can also be utilized during tokamak start-up to obtain desired magnetohydrodynamic equilibrium states~\cite{ebrahimi15,stanier13}.  

 \begin{figure}
\includegraphics[width=0.5\textwidth]{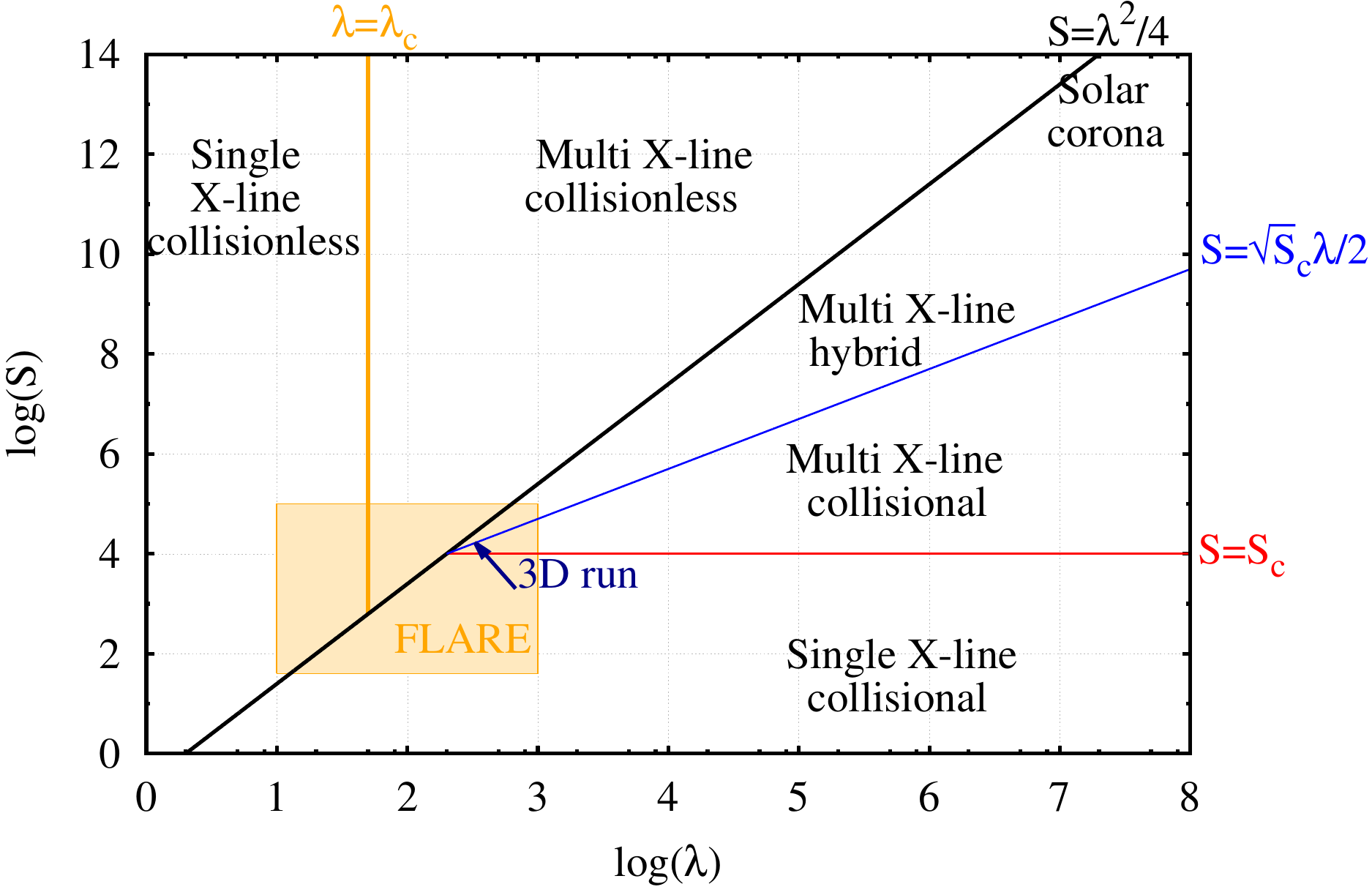}
\caption{\label{fig:phasediagram}Reconnection phase-diagram for Lundquist number $S$ and system-size $\lambda$ (see text for definitions). Different regimes of reconnection are delineated by the labelled (approximate) thresholds with $S_c=10^4$ and $\lambda_c=50$. The conditions for the solar corona and the FLARE experiment are marked. The tail of the arrow shows the initial conditions for the 3D kinetic simulation in this paper, and the head of the arrow gives the conditions just prior current-sheet break-up ($t\Omega_{ci}=88$).}
\end{figure}  

In these different plasma environments, the regimes of reconnection can vary, depending on the plasma size, collisionality, and the magnetic field configuration. Recent efforts~\cite{jidaughton11,baalrud11,daughtonroytershteyn12,cassak13,huangbhattacharjee13,le15,loureirouzdensky16,pucci17} have sought to classify the different regimes of reconnecting current sheets using a phase diagram in $S-\lambda$ space, for Lundquist number $S \equiv \mu_0\, v_{A}\, L_{CS}/\eta$ and normalized system-size $\lambda \equiv L/\delta_i$. Here, $v_{A}$ is the Alfv\'en velocity defined with the upstream (reconnecting) magnetic field, $L_{CS} = L/2$ is the current sheet half-length for a system of size $L$, $\eta$ is the Spitzer resistivity, and $\delta_i$ is the relevant ion kinetic scale. In a low-$\beta$ plasma, with $\beta$ the ratio of thermal to magnetic pressures, $\delta_i \equiv \rho_s = \sqrt{(T_i+T_e)m_i}/(q_iB)$ is the ion-sound radius defined with the ion/electron temperatures $T_{i/e}$, the magnetic field strength $B=|\boldsymbol{B}|$, and the ion charge $q_i$ and mass $m_i$. Figure~\ref{fig:phasediagram} shows an example phase-diagram that is similar to the one proposed in Ref.~[\onlinecite{jidaughton11}]. Here, the value of $S_c \sim 10^4$ is assumed to be the critical threshold at which a collisional Sweet-Parker current layer breaks up due to the plasmoid instability in MHD, although this can depend in practice on the background fluctuation level in the system~\cite{comisso16, huang17}. 
 
A long standing problem in reconnection theory has been a viable explanation for the fast ($S$-independent) reconnection rates in solar flares. The initially promising Petschek model~\cite{petschek64} invoked a microscopic value for the current sheet length, $L_{CS} \ll L$, with the primary energy conversion occuring at pairs of slow-mode shocks that bound the reconnection exhaust. However, an \textit{ad-hoc} localized resistivity enhancement is necessary to access the solution within resistive-MHD~\cite{ugai77,forbes13}, and it has not yet been validated with either first-principles numerical simulations or laboratory experiments (unlike the Sweet-Parker solution~\cite{ji98,daughton09b} with $L_{CS}\sim L$). 

An alternative idea invokes kinetic scales, following the now well established result from simulations~\cite{birn01} and experiments~\cite{yamada06,egedal07} that reconnection becomes fast when the Sweet-Parker current sheet thickness $\delta_{SP}=S^{-1/2}L_{CS}$ falls below the ion kinetic scale $\delta_i$. This transition was historically considered using laminar Sweet-Parker layers, e.g. Ref.~[\onlinecite{cassak06}], for which the threshold is the black line in Fig.~\ref{fig:phasediagram}. However, the Lundquist number in the corona~\cite{jidaughton11} $S\sim 10^{13}$ is vastly above $S_c$, and it is now widely recognised that current sheets will become unstable to the plasmoid instability before the laminar Sweet-Parker layers have time to form~\cite{puccivelli14,uzdenskyloureiro16,comisso16,huang17,pucci17}. Studies~\cite{bhattacharjee09,cassak09,huang10,loureiro12} have found that plasmoid-dominated reconnection can be fast in the ``Multi X-line collisional'' regime of Fig.~\ref{fig:phasediagram}, which can be modelled with resistive MHD simulations without invoking kinetic scales. However, the applicability of these results to solar flares remains uncertain for several reasons. 

Firstly, the onset of the plasmoid instability may lead to kinetic scale reconnection more readily than by the thinning of a laminar Sweet-Parker layer. This idea was first suggested by Ref.~[\onlinecite{shibata01}] who proposed that the plasmoid formation will lead to the formation of new secondary current sheets, which are also unstable to plasmoid formation. Applied recursively, this suggests a hierarchy of sheets and islands, which can form a cascade down to the ion kinetic scales where collisionless reconnection is triggered. This basic scenario has been confirmed in 2D using both Hall-MHD~\cite{shepherd10,huang11}, as well as fully kinetic simulations~\cite{daughton09a,daughton09b}, which give the theoretical basis for the blue line in Fig.~\ref{fig:phasediagram}. Within 3D reconnection layers, plasmoids are potentially unstable over a broader range of angles, and lead to the formation of flux ropes with considerably more freedom to interact. Large-scale 3D MHD simulations~\cite{oishi15,huang16,beresnyak17,kowal17} indicate that the reconnection layer becomes turbulent. While new thin current sheets are still produced, it is less clear how to estimate if this 3D dynamics leads to kinetic scale reconnection. 

Secondly, it is expected that the electric fields associated with solar flare reconnection should significantly exceed~\cite{cassakshay10,jidaughton11} the critical Dreicer~\cite{dreicer59} threshold, $E_{\textrm{flare}} \gg E_D = (m_e T_e)^{1/2}\nu_{ei}/e$, at which fluid models break down~\cite{daughton09a,roytershteyn10}. These super-Dreicer electric fields may play a role in the generation of non-thermal distributions of particles that are often observed during solar flares~\cite{lin06,krucker10}. 
      
The Facility for Laboratory Reconnection Experiments (FLARE~\cite{jiflare18}) has been designed to tackle these questions, amongst others. The maximum $S=10^4-10^5$ and $\lambda=10^2-10^3$ accessible are small compared with solar flare values, but should be large enough to study the phase transitions between the different reconnection regimes shown in Fig.~\ref{fig:phasediagram}. These more modest values are also becoming accessible for direct numerical simulation using first principles kinetic modelling, including the effects of Coulomb collisions~\cite{daughton09a,daughton09b,roytershteyn10}. In particular, Refs.~[\onlinecite{daughton09a,daughton09b}] have studied these phase-transitions with $2D$ simulations using the Harris sheet equilibrium in the $\beta \approx 1$ regime. At these lower values of $S$, the Sweet-Parker layer is able to form initially (in contrast to coronal values) but thins due to Joule heating along with a temperature dependent resistivity.  For small systems, reconnection transitions to the kinetic regime in laminar layers as $\delta_{SP}$ thins below $\delta_i$, but for larger layers this transition is triggered earlier by the onset of the plasmoid instability (as indicated by the blue line in Fig.~\ref{fig:phasediagram}). 

In the present paper, this transition is considered in 3D for an initially force-free current sheet in the low-$\beta$ regime, using a first-principles kinetic simulation with a Fokker-Planck collision operator. The low-$\beta$ regime is relevant for solar flares and magnetic reconnection experiments in FLARE. Compared with the $\beta \approx 1$ results of Ref.~[\onlinecite{daughton09a}], the low-$\beta$ current layer is found to thin much more rapidly from its initial thickness due to Joule heating and reach a significant Lundquist number $S\sim 10^4$ prior to plasmoid onset. At onset, the plasmoid instability in 3D results in multiple oblique modes that form at different rational surfaces~\cite{daughton11,baalrud12}, and can be stretched~\cite{huang17} and rotated by the reconnection outflow jets. 

It is found that the growth rates for the standard ($k_y \approx 0$) modes of the instability agree well with the semi-collisional predictions~\cite{drakelee77} of boundary layer theory for the tearing instability, but the angular cut-off for the unstable oblique modes is significantly smaller than predicted. Although the initial conditions are force-free, temperature gradients develop self-consistently due to Joule heating in the initial phase and the possibility of diamagnetic stabilization due to these gradients is considered. The temperature gradient stabilization predicted for the semi-collisional drift-tearing mode~\cite{connor12} is too small to explain this effect alone, but there may be additional stabilization due to the break-down of scale separation between the inner tearing layer thickness and the outer current sheet~\cite{baalrud18}. In the non-linear regime, the oblique tearing modes grow to form flux-ropes that undergo a variety of kink and coalescence processes, while they continue to be rotated by the reconnection outflows. These interactions lead to a turbulent-like state with large regions of stochastic magnetic field. 

Despite these complications, this simulation suggests that the transition from collisional to kinetic reconnection can occur in a manner analogous to the 2D picture. Thin current layers form between the flux-ropes, where super-Dreicer electric fields are supported by collisionless terms in Ohm's law. These thin current layers can become unstable to the generation of additional flux-ropes.

The paper is organized as follows. In Section~\ref{sec:setup}, the initial and boundary conditions, and the numerical parameters for the simulation are described. In Section~\ref{sec:results}, an overview of the different stages of the simulation is given. These stages are then considered in more detail in the following sections. Section~\ref{sec:thinning} describes the thinning of the collisional Sweet-Parker current layer prior to onset. Section~\ref{sec:obliqueplasmoid} presents an account of the oblique plasmoid instability and compares with current theories, with focus on the role of outflow jets in the stretching and rotation of the oblique modes, the collisionality of the inner tearing layer, the angular spectrum of the oblique modes, the non-linear flux-rope processes, and the generation of stochastic magnetic field. Section~\ref{sec:kinetictransition} describes evidence for the transition to kinetic reconnection in thin current layers that form due to the plasmoid instability. Finally,  a summary of results is given in Section~\ref{sec:conclude}.

\section{\label{sec:setup}Simulation set-up}

The primary simulation in this paper was performed with the VPIC particle in cell code~\cite{bowers08}. Unless otherwise specified, velocities are normalized by the light speed $c$, frequencies by the electron plasma frequency $\omega_{pe0}=\sqrt{n_0 e^2/\epsilon_0 m_e}$ and distances by the electron skin-depth $d_{e0} = c/\omega_{pe0}$. The simulation described in this paper is initialised with a force-free current sheet in a uniform plasma of physical number density $n_{i0}=n_{e0}=n_0$, and temperature $T_{i0}=T_{e0}=T_{0}$, with electrons (ions) of mass $m_e$ ($m_i$) and charge $-e$ ($e$). The initial magnetic field profile is given by
\begin{equation}\label{initialfield}\boldsymbol{B} = B_{r0} \tanh{\left(z/\delta_0\right)} \boldsymbol{\hat{x}} + B_{r0} \sqrt{b_g^2 + 1-\tanh^2{\left(z/\delta_0\right)}}\boldsymbol{\hat{y}},\end{equation}
where $B_{r0} = 1/(\omega_{pe0}/\Omega_{ce0})$ is the asymptotic reconnecting magnetic field, $b_g = 0.6$ is the ratio of the guide field to the reconnecting field, and $\delta_0 = 2\,d_{i0}$ is the initial current sheet half-thickness in units of the ion inertial length $d_{i0}/d_{e0} = \sqrt{m_i/m_e}$. The initial ratio of the electron thermal to magnetic pressure based upon the reconnecting field is $\beta_{e0} = 2 \mu_0 n_0 T_0/B_{r0}^2 = 0.08$, and the ratio of the electron plasma frequency to the gyro-frequency $\omega_{pe0}/\Omega_{ce0} = 1$ (similar to a solar coronal value). In order to start in a collisional (Sweet-Parker) parameter regime, a large seperation of scales is needed between the current sheet length and the ion kinetic scales ($\lambda > \sqrt{4S}$ according to Fig.~\ref{fig:phasediagram}). A reduced ion-to-electron mass ratio of $m_i/m_e = 40$ is used to make such simulations tractable.

The domain for the 3D simulation is a box of size $(L_x, L_y, L_z) =(164, 109.3, 54.7) \, d_{i0}$ that is periodic in $x$ and $y$, and has perfect conducting and particle reflecting boundaries in the $z$ direction. The spatial grid is $(n_x,n_y,n_z) = (3072, 2048, 1024)$ with $140$ particles per cell for each species (total $1.8 \times 10^{12}$ particles). The timestep is $\Delta t \omega_{pe} = 0.12$ (light wave CFL$=0.6$). 

Both ion and electron Coulomb collisions are studied. These are modelled using a Monte Carlo treatment of the Fokker-Planck collision operator~\citep{takizuka77,daughton09a}. The initial ratio of the electron-ion collision frequency to the cyclotron frequency is chosen to be $\nu_{ei0}/\Omega_{ce0} = 0.04$ such that the plasma is well magnetised. Since collisions are infrequent ($\omega_{pe0} = \Omega_{ce0} \gg \nu_{ei0}$), the collision operator is applied every $\Delta t_{\textrm{coll}} = 22\Delta t$ to reduce computational cost. This value was chosen based on numerical convergence to classical resistive friction~\cite{braginskii65} within the current sheet at early time - when the plasma is cold and the requirement to resolve the collision frequency is the most restrictive. 

The initial conditions described above can be understood in the context of the reconnection phase diagram (Fig.~\ref{fig:phasediagram}). For a low-$\beta$ force-free current sheet, the key parameters are the system-size $\lambda = L/\rho_{s}$, and the Lundquist number based on the parallel resistivity $S_\parallel=L_{CS}v_A\mu_0/\eta_\parallel$. The latter can be written as $S_\parallel = (L_{CS}/d_{i0})/\hat{\eta}_\parallel$ for normalized resistivity $\hat{\eta}_{\parallel} = 0.51 (\nu_{ei0}/\Omega_{ce0}) (T_0/T_e)^{3/2}$. We follow the conventions of Ref.~[\onlinecite{daughton09b}] to define $L=L_x$ and $L_{CS} = L_x/4$. At $t=0$, the initial conditions above give $\lambda_0 = 676$ and $S_{\parallel 0} = 2010$. This position is marked in Fig.~\ref{fig:phasediagram}(bottom) as the tail position of the blue arrow, which is within the `single X-line collisional' regime and the operating regime of the FLARE magnetic reconnection experiment. 

An additional requirement for collisional reconnection is for the electric field to be less than the Dreicer field $E_y/E_D < 1$. At early time (see below) the electric field is given by the resistive friction, $E_y = \eta_\parallel j_\parallel$, where $j_\parallel$ is the current at the X-point. It can be shown that 
\begin{equation}E_y/E_D = \frac{0.51}{\sqrt{\beta_e/2}(\delta/d_{i0})\sqrt{m_i/m_e}},\end{equation}
where $\beta_e$ is defined using the electron temperature and the upstream field. At $t=0$, $\beta_e=\beta_{e0}$ and $\delta=\delta_0$ to give $E_y/E_D=0.2$. 

A 2D perturbation is applied to the magnetic field to start the reconnection with $\boldsymbol{\delta B} = \boldsymbol{\nabla} \times \left(\delta A_y\boldsymbol{\hat{y}}\right)$, where
\begin{equation}\label{initialpert}\delta A_y = -\frac{0.0125 B_{r0} L_x}{\pi} \cos{\frac{2\pi (x-L_x/2)}{L_x}}\sin{\frac{\pi (z-L_z/2)}{L_z}}.\end{equation}

\section{\label{sec:results}Simulation results}

\begin{figure}
\includegraphics[width=0.48\textwidth]{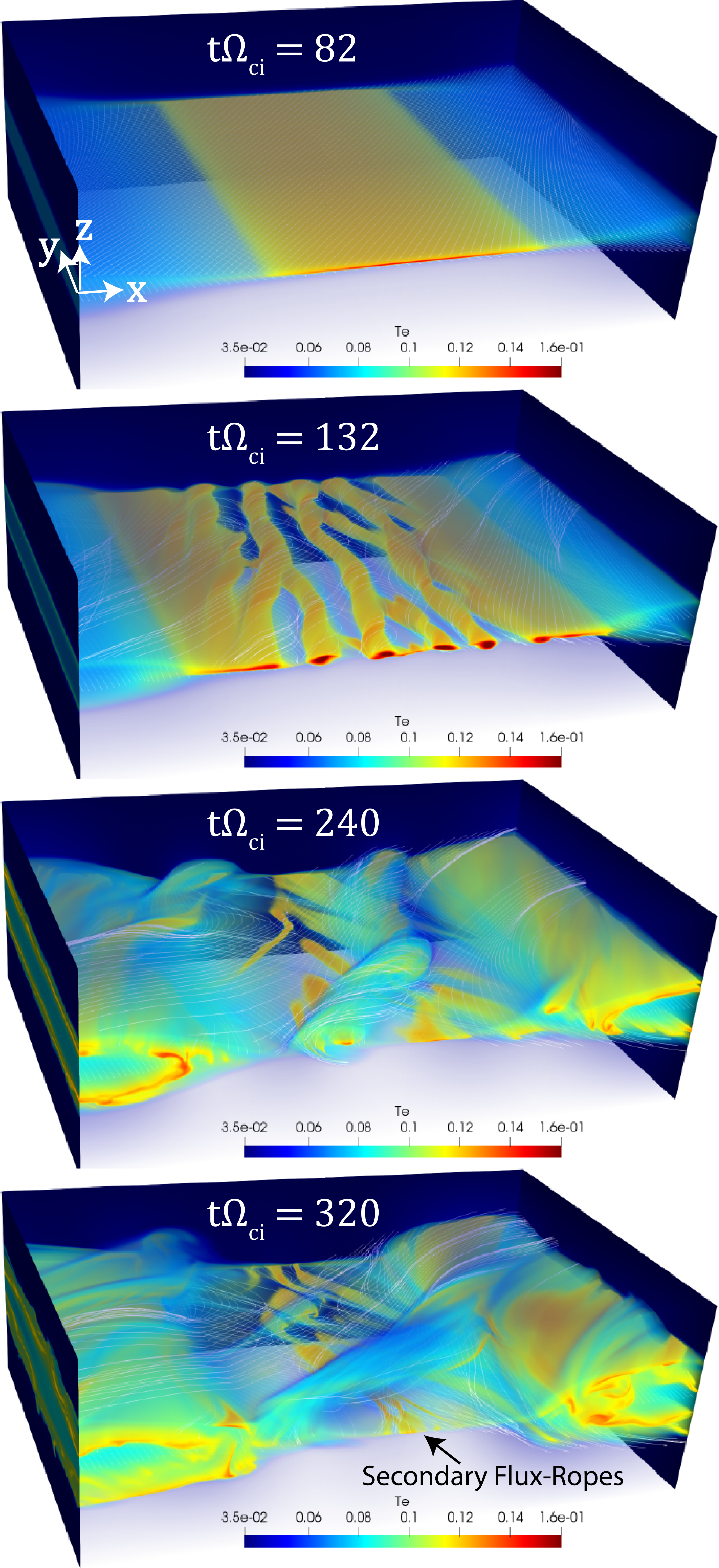}
\caption{\label{fig:snapshots}Volume rendering of the electron temperature $T_e$ with sample magnetic field lines (white). Movie (Multimedia view).}
\end{figure}
Figure~\ref{fig:snapshots} shows several snapshots of the electron temperature $T_e$ over the course of the simulation. Since electron heat transport is primarily along the magnetic field, $T_e$ serves as a useful proxy to visualise the magnetic topology. In the first snapshot, the $T_e$ profile is due to Joule heating within a quasi-2D current sheet structure that is set-up from the initial magnetic field perturbation. As will be discussed below, the electron heating leads to current layer thinning until the layer becomes unstable to the primary plasmoid instability.

The second panel shows this instability in the early non-linear phase. The formation of oblique flux-ropes breaks the initial symmetry, and they exhibit a range of kinking and coalescence processes. In the third panel, the magnetic flux-ropes produced by this instability are advected downstream, and further thin current layers form. These can also become unstable to secondary tearing-type instabilities to produce further flux-ropes as demonstrated in the fourth panel. The different stages of the simulation are discussed in further detail in the following sections.  

\section{\label{sec:thinning}Single X-line collisional reconnection}

\begin{figure}
\includegraphics[width=0.5\textwidth]{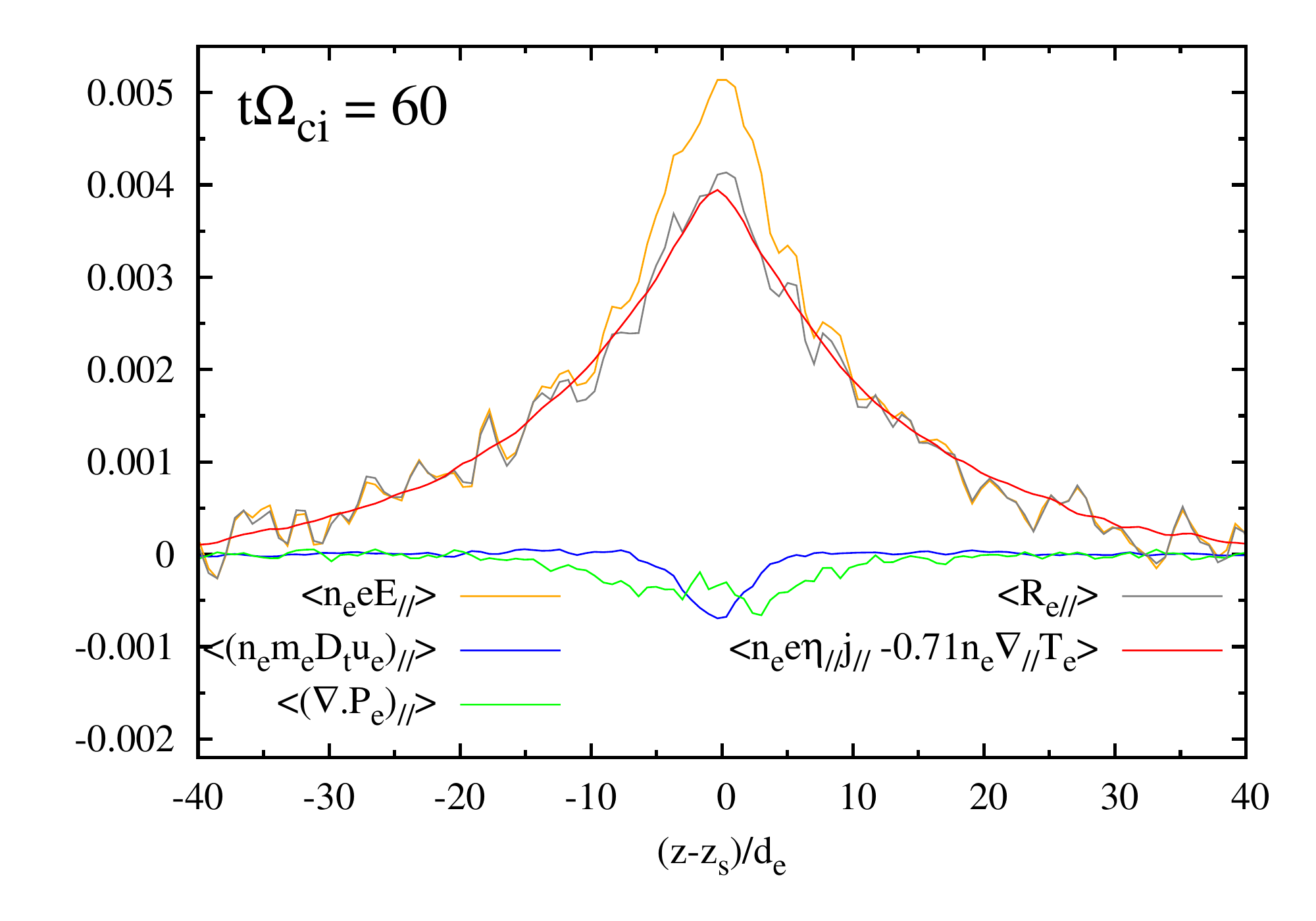}
\caption{\label{fig:ohmslaws}Contributions to parallel force balance at $t\Omega_{ci}=60$, in the initial phase of the simulation prior to plasmoid instability onset. Contributions have been averaged in time and along the field-lines - see text for definitions.  Quantities are expressed in ion units, after normalization by $n_0B_{r0}v_{A}$.}
\end{figure}

\subsection{\label{sec:SPlayer}Collisional current layer}

In order to verify that the current layer is in the collisional regime in the initial phase of the simulation, and to quantify the applicability of classical transport theory~\cite{braginskii65}, the parallel component of the electron momentum balance across the current sheet is considered. The parallel electron momentum equation (Ohm's law) is given by

\begin{equation}\label{ohmslaw}n_e e E_\parallel + \left[n_e m_e \frac{D \boldsymbol{u}_e}{D t}\right]_\parallel + \left[\boldsymbol{\nabla} \cdot \tensor{\boldsymbol{P}}_e\right]_\parallel = R_{e\parallel},\end{equation}
where $E_\parallel = \boldsymbol{E}\cdot \boldsymbol{\hat{b}}$, $\boldsymbol{\hat{b}}=\boldsymbol{B}/B$, $\boldsymbol{E}$ is the electric field, $\boldsymbol{u}_e$ is the electron bulk velocity, $D_t$ is the total derivative,  $\tensor{\boldsymbol{P}}_e$ is the electron pressure tensor, and $\boldsymbol{R}_e$ is the collisional momentum exchange, which is identically zero in a collisionless plasma. In the strongly magnetized and collisional regime, $R_{e\parallel}$ can be computed from classical transport theory~\cite{braginskii65} as 
\begin{equation}\label{neEfric}R_{e\parallel} \approx n_e e \eta_\parallel j_\parallel - 0.71 n_e \nabla_\parallel T_e,\end{equation}
where the first term is due to the resistive friction, and the second term is due to the parallel thermal force. To test the closure, all of the terms in Eqs.~(\ref{ohmslaw},\ref{neEfric}) were first averaged over a collision timescale $\nu_{ei}^{-1}$. Then, to further reduce statistical noise, the same terms were spatially averaged by integration along magnetic field-lines from an initial line of seed-points $\boldsymbol{x}=\boldsymbol{x}_0$ as
\begin{equation}<R_{e\parallel}> = \frac{1}{L_s}\int_{\boldsymbol{x}_0}^{\boldsymbol{x}_f} \boldsymbol{\hat{b}}[\boldsymbol{x}(s)]\cdot \boldsymbol{R}_e[\boldsymbol{x}(s)] ds,\end{equation}
where the final position $\boldsymbol{x}=\boldsymbol{x}_f$ is a distance $L_s$ along the field-line from $\boldsymbol{x}_0$. Ref.~[\onlinecite{le18}] has shown that this method of spatial averaging gives less smearing out of the diffusion regions compared to averaging along the $y$-axis when structures do not align with the $y$-direction, which is the case after onset of the oblique plasmoid instability.

Figure~\ref{fig:ohmslaws} shows the parallel force balance at $t\Omega_{ci} = 60$, when the reconnecting current layer has formed, but prior to plasmoid instability growth. Here, $\boldsymbol{x}_0 = (L_x/2,0,z)$ for $z\in [z_s-40,z_s+40]$, $z_s=L_z/2$, and $L_s=40$ ($d_e$). The term due to collisional momentum exchange, $<R_{e\parallel}>$  (grey), is calculated as the residual of the left hand side of Eq.~(\ref{ohmslaw}). It balances the term due to the parallel electric field, $<n_e e E_\parallel>$  (orange), and sets the thickness of the electron diffusion region at this time. The collisional transport closure (red) and the residual (grey) agree to within $3\%$ at the peak values, suggesting that classical transport is well founded in this early phase. Within the closure term, the friction term $<n_ee\eta_\parallel j_\parallel>$ is dominant over the thermal force $<-0.71 n_e \nabla_\parallel T_e>$. However, although the electron inertia (blue) and electron pressure tensor (green) are small, they are non-negligable in the center of the current sheet where they balance $20\%$ of the parallel electric field term at the X-point. A possible reason for this is partial runaway of electrons in the tail of the distribution function, which can occur even for sub-Dreicer electric fields~\cite{dreicer60,connorhastie75}.

\subsection{\label{sec:thinningmodel}Resistive thinning of current layer}

\begin{figure}
\includegraphics[width=0.5\textwidth]{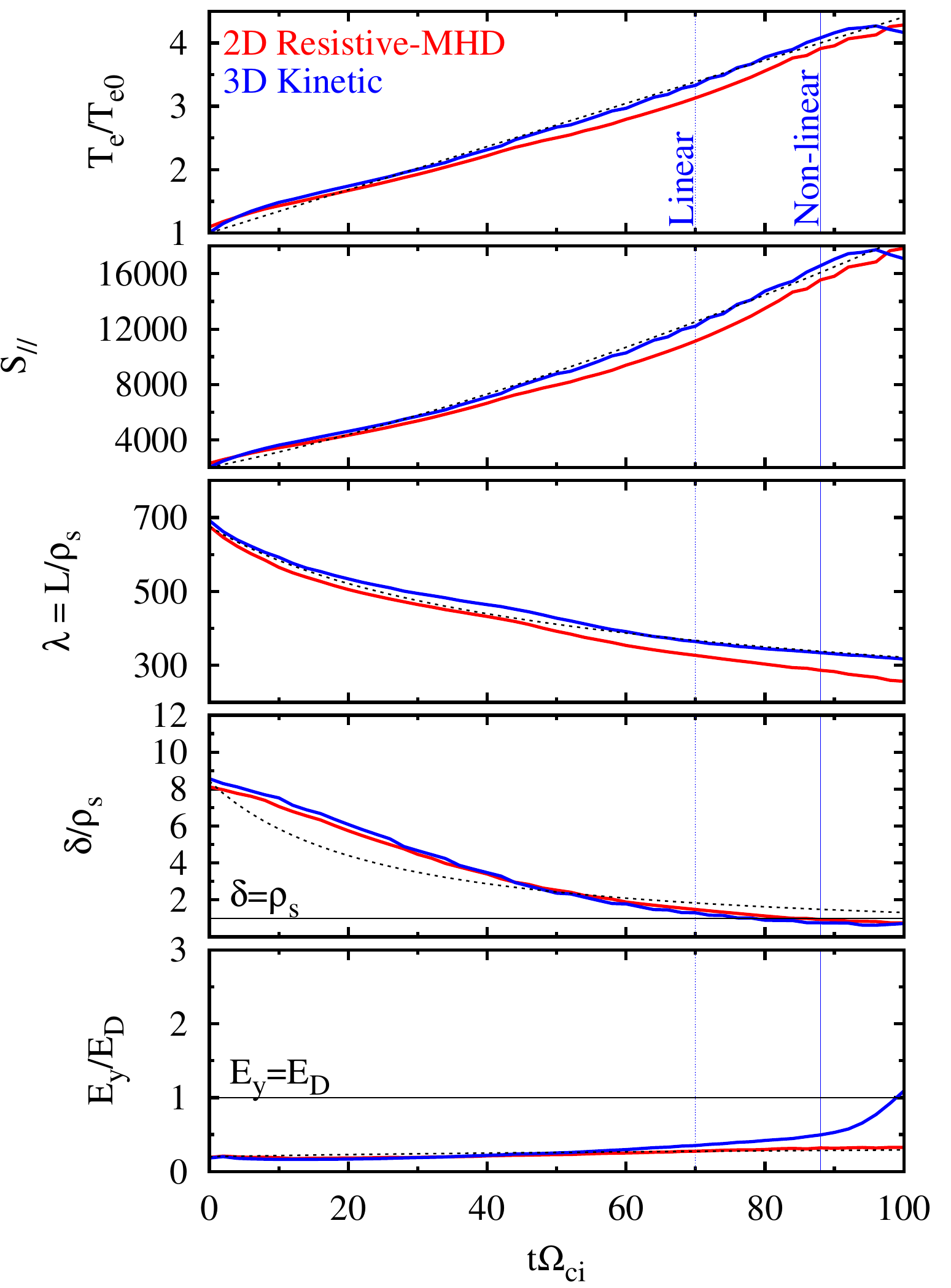}
\caption{\label{fig:thinning}Time traces of the electron temperature increase $T_e/T_{e0}$, the Lundquist number $S_\parallel$, the ratio of the system-size to the sound radius $\lambda=L/\rho_s$, the ratio of the layer thickness to the sound radius $\delta/\rho_s$, and the ratio of the electric field to the Dreicer runaway field $E_y/E_D$ in the early phase. Shown for the 3D first-principles kinetic simulation (blue) and a 2D single-fluid resistive MHD~\cite{chacon08b} simulation (red) assuming $T_i=T_e$ with a temperature dependant Spitzer resistivity. The black dashed lines show the simple scaling model discussed in the text. The blue vertical lines show the start of plasmoid instability growth (``Linear'' phase), and the time the magnetic islands are comparable to the current layer thickness (``Non-linear'' phase).}
\end{figure}

To characterize the initial current sheet thinning phase, prior to the plasmoid instability onset, Figure~\ref{fig:thinning} shows time traces of $T_e/T_{e0}$, $S_\parallel$, $\lambda=L/\rho_s$, $\delta/\rho_s$, and $E_y/E_D$ from the simulation (blue). Since the current sheet in the 3D simulation has symmetry along the $y$-direction (first panel in Fig.~\ref{fig:snapshots}), this 3D data is first reduced to 2D by averaging in $y$. The values of $T_e$, $T_i$, $|B|$, $n$ and $E_y = -\partial_t A_y$ are then measured at the dominant X-point of the mean-field magnetic flux profile, and $\delta$ is the half-thickness of the current layer at the thinnest point along its length (usually at the dominant X-point). The value of $\delta$ is estimated by fitting Eq.~(\ref{initialfield}), such that $\delta(t=0) = \delta_0$. 

For simplified fluid models with constant plasma resistivity, the current layer will thin due to the initial perturbation towards a constant Sweet-Parker thickness $\delta_{SP0} = S_0^{-1/2} L_{CS}$ for $S_0 \approx \textrm{const}$. Depending on the Lundquist number and the background noise level, the sheet may be either be stable, or break up before or after it is formed. In the present simulations, the plasma transport is determined self-consistently from the kinetic description of collisions and includes temperature dependent and anisotropic resistive and thermal friction, viscosity, heat conduction, and species thermal equilibration. Thus, $S_{\parallel} \neq \textrm{const.}$, and $\delta_{SP} = S_{\parallel}^{-1/2} L_{CS}$ can evolve in time. The precise evolution of the thickness $\delta(t)$ can, in principle, depend on all of the transport effects mentioned.

To illustrate the most important physics, the same parameters are computed from a 2D resistive MHD simulation (red) with corresponding initial conditions, a temperature dependent Spitzer resistivity~\cite{spitzer53}, and which neglects heat conduction and assumes exact temperature equilibration $T_i=T_e$. Here, for the single-fluid model, $\rho_s$ and $E_D$ are not physically meaningful, but are computed to normalize quantities in the same manner as the kinetic simulation. The simplified MHD model reproduces reasonably well the overall profiles $T_e/T_{e0}$, $S_\parallel$, $\lambda$ and $\delta/\rho_s$. The kinetic model has a slightly larger $T_e$ (and therefore $S_\parallel$) than the MHD model, which is attributed to the preferential Joule heating of electrons while the equilibration timescale $\tau_{eq}\Omega_{ci} = (T_e/T_{e0})^{3/2}/(\nu_{ei0}/\Omega_{ce0}) = 25 (T_e/T_{e0})^{3/2}$ does not remain small compared to the timescale of current layer thinning. Despite this, the temperature ratio remains within a factor of $\tau \equiv T_e/T_i = 1.5$ at $t\Omega_{ci}=60$, and $\tau = 2$ at $t\Omega_{ci}=90$. Other noticeable differences include a slightly larger~\footnote{$(T_i+T_e)$ is $20\%$ larger in MHD at $t\Omega_{ci}=60$, presumably due to neglecting heat conduction} total temperature and thus $\rho_s$ in the MHD model, and a significantly weaker $E_y/E_D$ at late times. To verify that the thinning observed requires the temperature dependent resistivity, we performed a similar resistive-MHD simulation with uniform resistivity and found that $\delta$ is approximately $3$ times thicker (not shown) at $t\Omega_{ci}=90$. This result demonstrates that the temperature dependent (Spitzer) resistivity can play an important role in the evolution towards plasmoid unstable regimes.

With the results described above, it is convenient to parameterize the thinning via a simplified analytic scaling model that can be used to plot the trajectory of the thinning phase onto the reconnection phase diagram. Firstly, based on the simulation data, it is assumed that  $n$, $|B|$ and $L_{CS}$ are constant, and that $T_i   \approx T_e$. With these assumptions, the phase-diagram co-ordinates vary only with $T_e/T_{e0}$ as $S_\parallel \propto \eta_\parallel^{-1} \propto (T_e/T_{e0})^{3/2}$ and $\lambda \propto \rho_s^{-1} \propto (T_{e}/T_{e0})^{-1/2}$. Then the temperature evolution can be estimated by neglecting heat conduction and viscous heating (which occurs primarily downstream of the X-point), such that the temperatures increase solely due to Ohmic heating within the layer 
\begin{equation}\frac{3}{2}n_0 \partial_t (T_e + T_i) \approx \eta_{\parallel} j_{\parallel}^2.\end{equation}
Finally, it is assumed that the current at the X-point follows a Sweet-Parker scaling $j_{\parallel} \propto \delta_{SP}^{-1} \propto \eta_{\parallel}^{-1/2}$, such that $\partial_t T_e \propto \eta_{\parallel}^0$, i.e. an electron temperature that increases linearly in time $T_e \propto t$. The fractional heating rate can be estimated based upon the initial current density~\cite{daughton09a}, as $T_{e}/T_{e0} \approx 1 + Q_e t\Omega_{ci}$ where

\begin{equation}\label{heatingrate}Q_e \approx \frac{4}{6\, S_{\delta 0}\, \beta_{e0} (\delta_0/d_{i0})} \approx 0.0425,\end{equation}
with $S_{\delta 0} = S_{\parallel 0} \delta_0/L_{CS}$. It follows that $S_\parallel \approx S_{\parallel 0} (1 + Q_e t\Omega_{ci})^{3/2}$, $\lambda \approx \lambda_0 (1 + Q_e t\Omega_{ci})^{-1/2}$, $\delta \approx \delta_0 (1 + Q_e t\Omega_{ci})^{-3/4}$ ($\delta/\rho_s \approx \delta_0/\rho_0 (1 + Q_e t\Omega_{ci})^{-5/4}$), and $E_y/E_D \propto (1 + Q_e t\Omega_{ci})^{1/4}$. 

To compare the simple model against the simulation data, a linear profile (black dashed line) is fit to $T_e/T_{e0}$ for the kinetic simulation, which gives a measured value of $Q_e = 0.034$. The dashed lines in the other panels show predicted time profiles for each quantity using this measured value of $Q_e$, which give reasonable overall agreement with the data considering the number of assumptions made. Departures from these scalings are most noticeable in $\delta/\rho_s$ at early time, as it takes some time for Sweet-Parker reconnection to develop from the initial perturbation, and in $E_y/E_D$ at late time where $E_y$ deviates from $\eta_{\parallel} j_{\parallel}$ due to finite contributions from the pressure tensor and inertial terms in the momentum balance as discused above. These terms, which become significant during the early non-linear phase of the plasmoid instability (see Section~\ref{sec:kinetictransition}), are not present in the MHD model.

The peak values of $T_e/T_{e0}$ and $S_\parallel$ are approximately $2.5$ and $6$ times larger respectively than simulations with similar parameters\footnote{Here $T_e/T_{e0} \approx 4$ and  $S_\parallel=1.6\times 10^4$ at $t\Omega_{ci} = 88$, compared with $T_e/T_{e0}\approx 1.6$ for a simulation with $\eta_\perp = 0.04$, $\delta_0/d_{i0}=1$, and $L_x=100d_i$ in Ref.~[\onlinecite{daughton09a}], and $S_{\textrm{max}} = 2500$ for a simulation with $\eta_\perp = 0.04$, $\delta_0/d_{i0}=2$, and $L_x=200d_i$ in Ref.~[\onlinecite{daughton09b}].} reported in Refs.~[\onlinecite{daughton09a}] and [\onlinecite{daughton09b}] for the Harris sheet with $\beta \approx 1$. This follows from Eq.~(\ref{heatingrate}), where the fractional heating rate increases as $Q_e \propto \beta_{e0}^{-1}$ with other quantities equal.

\section{\label{sec:obliqueplasmoid}Oblique plasmoid instability}

\begin{figure}
\includegraphics[width=0.5\textwidth]{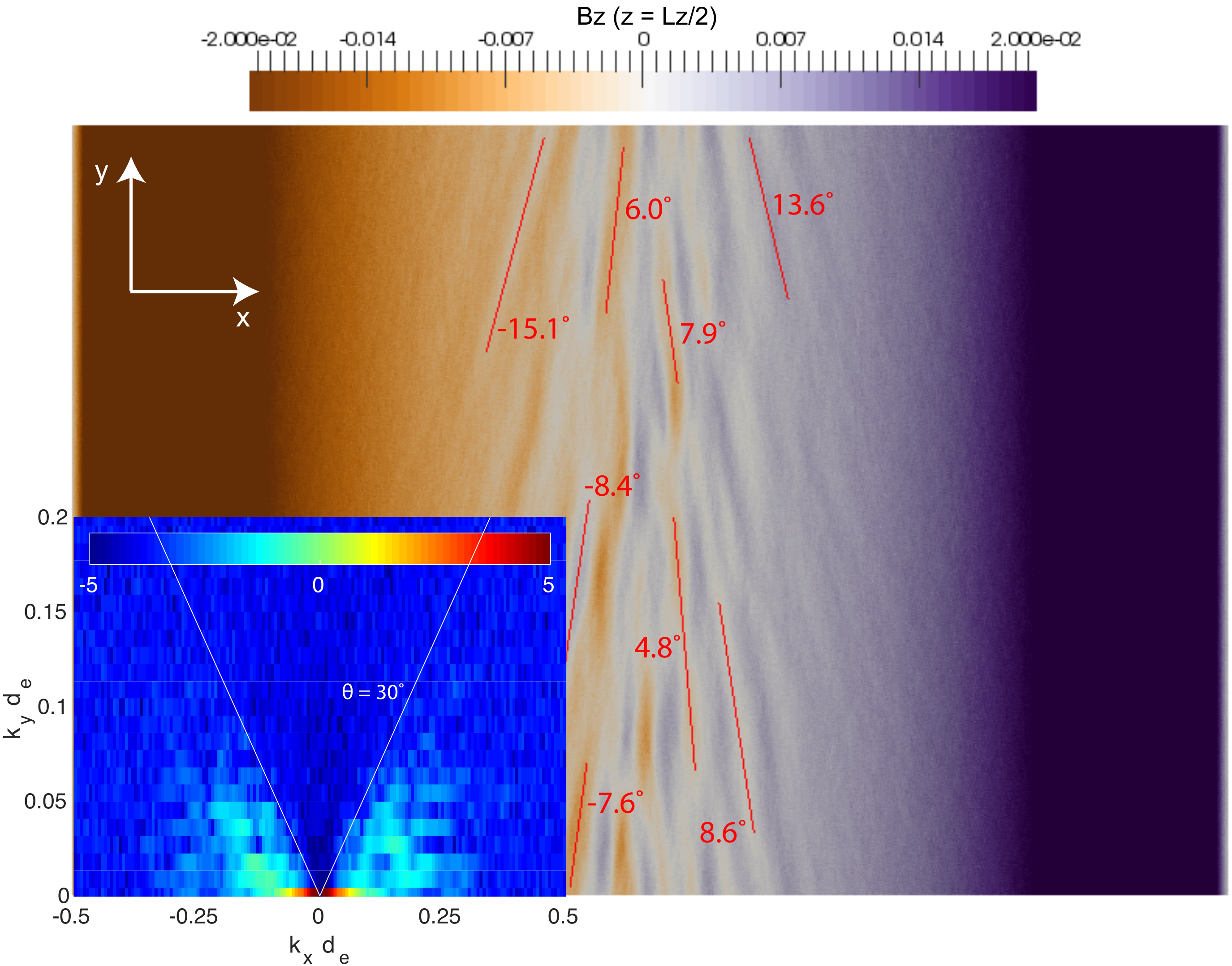}
\caption{\label{fig:obliquespectrum}Reconnected component of the magnetic field $B_z(x,y,z=L_z/2)$ at $t\Omega_{ci0} = 88$. The angles of tearing fluctuations with respect to the $y$-axis are indicated by labelled red lines. Inset: 2D Fourier spectrum, $P(k_x,k_y,t\Omega_{ci}=88)$. The labelled white lines mark angles $\theta = \pm 30^\circ$. The time evolution of this Fourier spectrum is available in the multimedia view (Multimedia view).  }
\end{figure}

Figure~\ref{fig:obliquespectrum} shows  $B_z$, the reconnected component of the magnetic field, in a top down view of the $z=0$ plane at $t\Omega_{ci0}=88$. At this time, which is indicated by the second vertical blue line in Fig.~\ref{fig:thinning}, tearing-type fluctuations in the current density become noticeable over the background current sheet structure. These fluctuations are visible in Fig.~\ref{fig:obliquespectrum} close to the center of the current layer, where they form at a range of oblique angles to the $y$-axis. To more clearly show the angular distribution of the fluctuations, Fig.~\ref{fig:obliquespectrum} inset shows $P(k_x,k_y,t\Omega_{ci}=88) = \log_{10}{\left(\int_0^{L_z} |\hat{B_z}(k_x, k_y,z)|^2\,dz\right)}$, the power spectrum of the magnetic energy density in $k_x-k_y$ space and integrated over the height of the simulation box $L_z$. Here, the peak values at $k_y = 0$ and $k_x d_e < 0.1$ are partly associated with the background reconnecting current sheet structure, but there is significant power across a range of oblique modes with $\theta = \arctan{(k_y/k_x)} \lesssim 30^\circ$. 

A full analysis of the plasmoid instability requires accounting for the detailed plasma physics of the inner tearing layer~\cite{fkr63,coppi79,drakelee77,cowley86}, the evolution of the background current profiles during the current sheet thinning process~\cite{uzdenskyloureiro16,puccivelli14,comisso16} ($\delta = \delta(t)$), and the role of outflow jets in the advection and stretching of weakly growing modes~\cite{huang17}. The full analysis is not given here, but the relative importance of each of these is examined in this section from the simulation data in comparison with current theories. In particular, we quantify the importance of plasma collisions in the inner tearing layer and investigate the physics responsible for the maximum cutoff angle $\theta_{\textrm{cutoff}} \approx 30^\circ$ observed.  

\subsection{Mode stretching and rotation by outflow jets}

\begin{figure}
\includegraphics[width=0.5\textwidth]{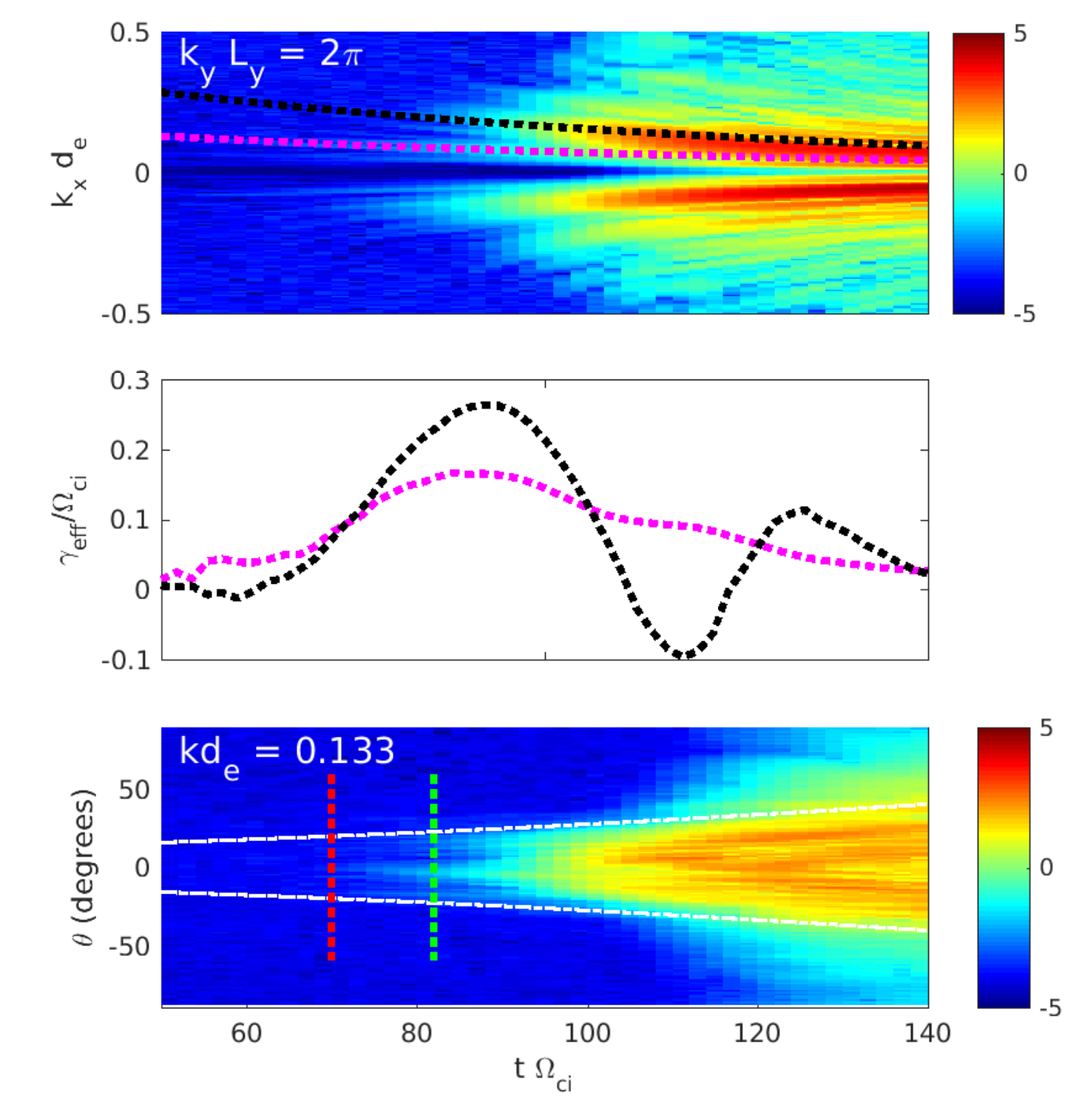}
\caption{\label{fig:kxthetatime} Top panel: Power spectrum of modes $P(k_x, k_yL_y=2\pi, t)$ in the $k_x-t$ plane for $t\Omega_{ci} \in [50, 140]$. The black and magenta dotted lines show two sample trajectories $k_x(t)$ from Eq.~(\ref{stretching}). Middle panel: Growth rates $\gamma_{\textrm{eff}}/\Omega_{ci}$ along the two characteristic trajectories calculated as in Eq.~(\ref{growthdef}). Bottom panel: Power spectrum of modes $P(k_x,k_y,t)$ in the $\theta-t$ plane for constant $kd_e=0.133$. The white dash-dotted lines show sample trajectories from Eq.~(\ref{rotation}), and the vertical dashed red and green lines show sample times at which the growth rates are compared with boundary layer theory in Fig.~\ref{fig:growthrates}.}
\end{figure}

The multimedia view of Fig.~\ref{fig:obliquespectrum} shows the time evolution of the power spectrum $P(k_x,k_y,t)$, with frames every $2 \Omega_{ci}^{-1}$ from $t\Omega_{ci}=0$ to $t\Omega_{ci}=140$. As well as the growth of the oblique modes, there is notable advection of these modes towards $k_x = 0$ due to mode stretching by the reconnection outflow jets. Fig.~\ref{fig:kxthetatime} (top panel) shows a slice of the power spectrum in the $k_x-t$ plane for $t\Omega_{ci}\in[50,140]$ at constant $k_y L_y = 2\pi$ ($k_yd_e=0.0091$), where the background 2D current sheet profile with $k_y=0$ is not visible. The oblique modes are initially visible at $t\Omega_{ci}=70$ where they are slowly advected towards $k_x = 0$. Ref.~[\onlinecite{huang17}] has studied this effect in detail with 2D resistive MHD simulations (without oblique modes), and generalized a model of the plasmoid instability in time evolving current sheets by Ref.~[\onlinecite{comisso16}] to account for this physics. In the model, the modes are assumed to be advected in the $k_x$-direction as $d_t k_x = -k_x v_x^\prime$ such that they follow characteristic trajectories
\begin{equation}\label{stretching}k_x = k_{x0} e^{-v_x^\prime t}.\end{equation}
Here, $k_{x0}$ is the initial component of the wavenumber in the $x$-direction, and $v_x^\prime$ is the gradient of the outflow jet velocity $v_x^\prime \approx v_{x,\textrm{max}}/L_{CS}$ for maximum outflow velocity $v_{x,\textrm{max}}$ and current sheet length $L_{CS}$. Two of these trajectories are plotted as the magenta and black curves in Fig.~\ref{fig:kxthetatime} (top panel), where $v_{x,\textrm{max}} \approx 0.5 v_A$ and $L_{CS} \approx L_x/4 = 41 d_i$ are assumed constant in time. The curves follow the visible mode stretching reasonably well. 

Since there are no outflow jets in the $y$-direction, the modes remain with approximately constant $k_y \approx k_{y0}$ (multimedia view of Fig.~\ref{fig:obliquespectrum}). An interesting consequence of this in 3D is that oblique modes rotate towards larger oblique angles due to the shear of the outflow jets. Using Eq.~(\ref{stretching}) for $k_x$ in the definition of the oblique angle $\theta = \arctan{(k_y/k_x)}$ gives the rotation as
\begin{equation}\label{rotation}\theta = \arctan{[\tan{(\theta_0)} \exp{(v_x^\prime t)}]},\end{equation}
where $\theta_0 = \arctan{(k_{y0}/k_{x0})}$. Fig.~\ref{fig:kxthetatime} (bottom panel) shows a slice of the mode spectrum in the $\theta-t$ plane for constant $kd_e = 0.133$. There is a slow but observable advection towards larger $|\theta|$, where the white dash-dotted lines show two sample trajectories in $\theta-t$ from Eq.~(\ref{rotation}).

\subsection{Collisionality of inner tearing layer}

The role of collisionality in the inner tearing layer depends on the relative magnitudes of the mode frequency $|\omega_r + i \gamma|$ and the collision frequency $\nu_{ei}$. The real frequency $\omega_r$ can be non-zero in the presence of temperature or density gradients across the rational surface, and we will discuss this further below. Following Ref.~[\onlinecite{huang17}], the power in each fourier mode can be modeled as
\begin{equation}\frac{d |\hat{B_z}(k_x,k_y,t)|^2}{dt} = \left[2\gamma(t) - v_x^\prime \right] |\hat{B_z}(k_x,k_y,t)|^2,\end{equation}
where $d_t = \partial_t - k_x v_x^\prime \partial_{k_x}$ is the derivative along the characteristics, and the growth rate $\gamma(t) = \gamma(\delta(t), k_x(t))$ depends upon the instantaneous current sheet thickness $\delta(t)$~\cite{comisso16}. Modes only grow when the growth-rate is large enough to overcome the mode stretching~\cite{huang17}, $\gamma(t) > v_x^\prime/2$. Rearranging this for the growth rate gives
\begin{equation}\label{growthdef}\gamma(t) = \frac{1}{2}\frac{d\left( \ln{|\hat{B_z}|^2}\right)}{dt} + v_x^\prime/2 \equiv \gamma_{\textrm{eff}}(t) + v_x^\prime/2.\end{equation}
Fig.~\ref{fig:kxthetatime} (middle panel) shows $\gamma_{\textrm{eff}}(t)/\Omega_{ci}$ calculated along the two curves in $k_x-t$ from the top panel using the data $P(k_x, k_yL_y=2\pi, t)$. Here, we have filtered the signal to remove high frequency waves while well preserving the time (zero phase delay) and peak magnitude of $\gamma_{\textrm{eff}}$. At $t\Omega_{ci} = 70$ both curves have $\gamma_{\textrm{eff}}/\Omega_{ci} \approx 0.05$, which is already significantly larger than $v_x^\prime/2\Omega_{ci} \approx 0.006$. At and after this time, the mode stretching is not a substantial effect and is neglected in the rest of the discussion on the linear growth with the assumption that $\gamma(t) \approx \gamma_{\textrm{eff}}(t)$. 

To estimate the importance of collisions, $\gamma/\Omega_{ci}$ can be compared with the electron-ion collision frequency $\nu_{ei}/\Omega_{ci} \approx (\nu_{ei}^0/\Omega_{ce}^0) (m_i/m_e) (T_e/T_{e0})^{-3/2} = 1.6 \,(T_e/T_{e0})^{-3/2}$. At $t\Omega_{ci} = 70$, $T_e/T_{e0} = 3.3$ (Fig.~\ref{fig:thinning}) such that $\nu_{ei}/\Omega_{ci} \approx 0.26$ is approximately $5$ times larger than $\gamma/\Omega_{ci}$ in Fig.~\ref{fig:kxthetatime} at this time. At a later time $t\Omega_{ci} = 82$, $T_e/T_{e0} = 3.8$ and $\nu_{ei}/\Omega_{ci} \approx 0.21$ is comparable to the instantaneous $\gamma/\Omega_{ci}$ of the two curves. We now proceed to compare the measured growth rates with those predicted from linear boundary layer theory in more detail.  

\subsection{Comparison with semi-collisional theory}

Depending on the plasma collisionality, different asymptotic regimes of the tearing instability have been derived in the literature. In the collisionless (CLS) regime, electrons within a channel of thickness $\Delta_\textrm{CLS}$ from the rational surface $z_s$ ($|z -z_s| < \Delta_\textrm{CLS}$) are freely accelerated along the field-lines by the induced electric field of the mode. For $|z-z_s| \geq \Delta_\textrm{CLS}$ the Doppler frequency becomes larger than the mode frequency, $\omega_d \equiv k_{\parallel} v_{Te} \geq |\omega|$, and the electrons experience an alternating electric field that significantly reduces the current response. The thickness of the channel $\Delta_\textrm{CLS}$ is found at $|z-z_s|=\Delta_\textrm{CLS}$ where $|\omega|=\omega_d$. Using $k_\parallel \approx k(z-z_s)/L_s$, for a magnetic shear length $L_s$ (defined below), gives
\begin{equation}\Delta_{\textrm{CLS}} = \frac{|\omega| L_s}{k v_{Te}}.\end{equation}
Ref.~[\onlinecite{drakelee77}] derive a growth rate for this regime, under the assumption of cold ions, as 
\begin{equation}\label{collisionlessnormal}\gamma_{CLS} = \frac{kv_{Te} d_e^2 \Delta^\prime}{2\sqrt{\pi} L_s}.\end{equation}
Here $\Delta^\prime$ is the parameter used to match asymptotic solutions from the outer ideal region $|z-z_s| \sim \delta \gg \Delta_\textrm{CLS}$ to the inner region $|z-z_s| \sim \Delta_\textrm{CLS}$. $\Delta^\prime$ is assumed small in the derivation of Eq.~(\ref{collisionlessnormal}). 

In this Section, it is assumed that the outer region is described by a 1D force-free profile. This is not strictly true for $t>0$, as reconnected ($B_z$) field develops within the current sheet during the initial Sweet-Parker phase giving a weakly 2D profile~\cite{loureiro13}, and the profile deviates from a force-free one due to Joule heating. Despite this, we find that profiles of the form of Eq.~(\ref{initialfield}) fit reasonably well the magnetic field data at $x=L_x/2$ for a fitting parameter $\delta(t)$. We thus consider below the role of temperature gradients only on the inner region. Eq.~(\ref{initialfield}) gives~\cite{baalrud12,liu13,akcay16} 
\begin{equation}\label{deltaprime}\Delta^\prime = \frac{2}{\delta}\left(\frac{1}{k\delta}\left(1+b_g^2 \tan^2{\theta}\right) - k\delta\right),\end{equation}
\begin{equation}\label{shearlength}L_s = \frac{k}{k_\parallel^\prime(z_s)} = \frac{\delta \sqrt{1+b_g^2}}{\cos{\theta} \left(1-b_g^2 \tan^2{\theta}\right)},\end{equation}
and
\begin{equation}\label{rationalsurface}z_s = - \delta \arctanh{\left(\sqrt{1+b_g^2} \sin{(\theta)}\right)}.\end{equation}

As discussed above, $\nu_{ei} \geq \gamma$ for the early phase of the instability, and thus it is necessary to include the effects of collisions. In the semi-collisional regime ($\nu_{ei} \gg |\omega|$, $\Delta_{SC} \ll \rho_s$), the thickness of the current channel $|z-z_s| = \Delta_{SC}$ is found when the mode frequency is balanced by collisional diffusion of electrons along field-lines~\cite{drakelee77}, $|\omega| = k_\parallel^2 v_{Te}^2/\nu_{ei}$. The inner layer is thus broadened by collisions as 
\begin{equation}\label{deltasc}\Delta_{SC} = \Delta_{\textrm{CLS}} (\nu_{ei}/|\omega|)^{1/2}.\end{equation}
Ref.~[\onlinecite{drakelee77}] has also derived closed form expressions for the growth rate in this semi-collisional regime under the assumptions of cold ions, small $\Delta^\prime$, and for weak density and temperature gradients. The growth rate is modified as\footnote{In Alfv\'en units~\cite{zocco11}, this is the same small-$\Delta^\prime$ growth rate used for a recent model of the semi-collisional plasmoid instability~\cite{bhat18}: $\gamma_{SC}/\omega_A \sim (kL_S)^{2/3} S^{-1/3} (\Delta^\prime \rho_s)^{2/3}$, where $\omega_A = L_s/v_A$, $S=L_s v_A/\eta$, and $\eta = d_e^2\nu_{ei}$.} 
\begin{equation}\label{semicollisionalnormal}\gamma_{SC} = \left[\frac{3 \pi^{1/4}}{4 \Gamma(11/4)}\right]^{2/3} \gamma_{CLS}^{2/3} \,\nu_{ei}^{1/3}.\end{equation}

\begin{figure}
\includegraphics[width=0.5\textwidth]{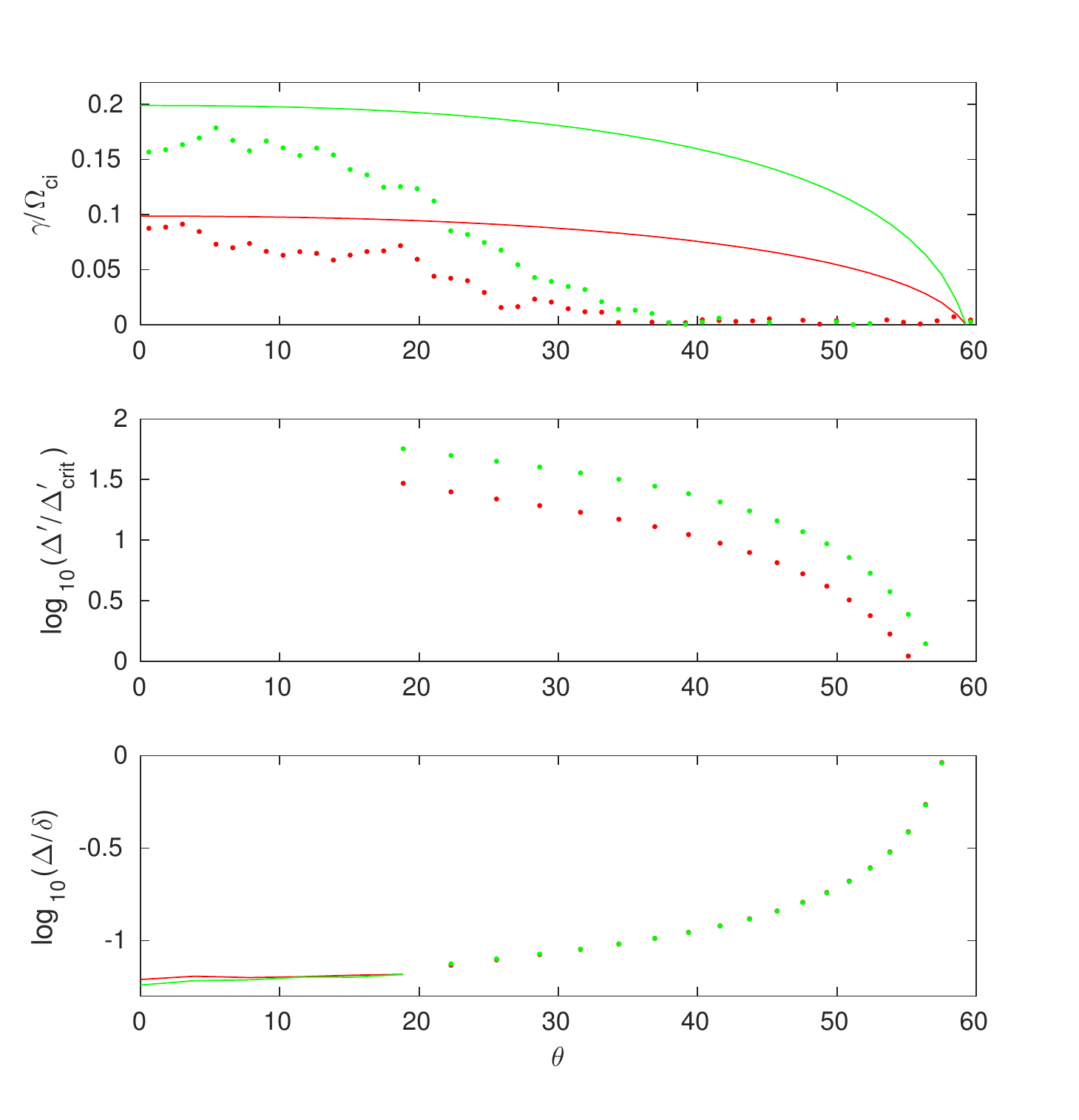}
\caption{\label{fig:growthrates} Top panel: Dotted lines show measured growth rates $\gamma_{\textrm{eff}}(\theta)/\Omega_{ci}$ at fixed $kd_e=0.133$ for $t\Omega_{ci}=70$ (red) and $t\Omega_{ci}=82$ (green). Solid lines show the predictions for the semi-collisional tearing mode from Eq.~(\ref{semicollisionalnormal}). Middle panel: The asymptotic matching parameter $\Delta^\prime$ from Eq.~(\ref{deltaprime}) divided by the critical value for marginal stability from boundary layer theory, $\Delta^\prime_{\textrm{crit}}$ in Eq.~(\ref{deltaprimecrit}), for the drift tearing mode on a logarithmic scale. Bottom panel: The ratio of the inner layer thickness $\Delta$ to the outer ideal region thickness $\delta$ on a logarithmic scale. Here $\Delta$ is calculated from Eqs.~(\ref{deltasc}) for modes with $\theta < 20^\circ$ (solid lines) and with Eq.~(\ref{deltaT}) for modes with $\theta>20^\circ$ (dotted lines).}
\end{figure}

Figure~\ref{fig:growthrates} (top panel) shows the measured growth-rates $\gamma(\theta)/\Omega_{ci}$ ($\gamma = \gamma_{\textrm{eff}}$) for fixed $kd_e = 0.133$ at $t\Omega_{ci} = 70$ (red dots) and $t\Omega_{ci} = 82$ (green dots), where these times are indicated by vertical lines in Fig.~\ref{fig:kxthetatime} (bottom panel). These are not the fastest growing modes in the simulation, but for $kd_e=0.133$ the small-$\Delta^\prime$ theory is appropriate ($\Delta^\prime \rho_s^{1/2} \Delta_{SC}^{1/2} < 1$)~\cite{zocco11}. In addition to the time filtering mentioned above, we have taken the mean of the positive and negative $\theta$ values to better compare with theory. The solid lines show the predicted growth rate $\gamma_{SC}$ from Eq.~(\ref{semicollisionalnormal}), where the collision term $\nu_{ei}$ is evaluated based on the local electron temperature $T_e(z)$ at the rational surface $z=z_s(\theta)$. 

Despite the assumptions that have been made in Eq.~(\ref{deltaprime}-\ref{semicollisionalnormal}), namely that the profile remains a 1D force-free layer with cold ions, there is fairly good agreement for $\theta < 20^{\circ}$ between the measured growth rates and Eq.~(\ref{semicollisionalnormal}). It should also be noted that $\gamma/\nu_{ei} \sim 1$ at $t\Omega_{ci} = 82$ which is not strictly in the regime of validity for the semi-collsional mode ($\nu_{ei} \gg \gamma$). 

Refs.~[\onlinecite{loureirouzdensky16,bhat18}] have argued that the onset of the plasmoid instability can occur earlier in the semi-collisional regime ($\Delta_{SC} \ll \rho_s$) than the resistive-MHD regime ($\Delta \gg \rho_s$), due to faster tearing mode growth rates. The precise threshold for onset in the semi-collisional regime is not considered here, but we note that onset occurs at a later time ($t\Omega_{ci} \approx 140$) in the 2D resistive-MHD simulation of Fig.~\ref{fig:thinning} than the kinetic simulation, despite the addition of a continuous random noise forcing term to the MHD velocity fields with amplitude larger than the PIC simulation noise level. 

\subsection{Stabilization of oblique modes}

Although there is good agreement for the modes with $\theta < 20^{\circ}$, there is clear disagreement between Eq.~(\ref{semicollisionalnormal}) and the measured growth rates for $\theta \gtrsim 20^{\circ}$. For $b_g = 0.6$, Eq.~(\ref{semicollisionalnormal}) predicts stabilization for $\theta = 59^{\circ}$ when $L_s \rightarrow \infty$ and $z_s\rightarrow \pm \infty$. The angle $\theta = 59^\circ$ is simply half the shear angle the magnetic field makes as it rotates across the current sheet, and is thus set by the background magnetic profile of the outer region. In contrast, the measured growth-rates are stabilized for $\theta \gtrsim 35^\circ$, at which the rational surface $z_s = \pm 0.81\delta$ is still within the current layer. This suggests there is some additional stabilization mechanism associated with the inner region. A possible explanation for the discrepancy, which will be presently considered, is the diamagnetic stabilization of oblique modes due to temperature and/or density gradients~\cite{drake83,cowley86,baalrud18}. Such diamagnetic flows do not exist in the force-free initial conditions, but become finite over time. Here we consider only electron temperature gradients, which arise mainly due to the Joule heating, as we find density gradients and ion temperature gradients to be significantly smaller.  

The diamagnetic frequency due to gradients in $T_e$ is given by $\omega_T = k T_e/(eBL_T)$, where $L_T = 1/|\partial_z \ln{(T_e)}|$. For the standard tearing modes ($\theta = 0$), $\partial_z T_e \approx 0$ due to the symmetry of the current layer, but $\omega_T \neq 0$ for oblique modes. The marginal stability threshold for standard tearing modes ($\Delta^\prime \geq 0$) is then increased to $\Delta^\prime \geq \Delta^\prime_{\textrm{crit}}$ for both collisionless~\cite{coppi79,antonsen81,cowley86} and semi-collisional~\cite{drake83,cowley86,connor12} drift-tearing modes. For the semi-collisional case, Ref.~[\onlinecite{drake83}] found $\Delta^\prime_{\textrm{crit}} \sim \hat{\beta}_T/\rho_{se}$ for the cold ion limit where $\beta_T = (\beta_e/2) L_s^2/L_T^2$. Ref.~[\onlinecite{cowley86}] then generalized this to include the effects of finite ion orbits in the regime with $\Delta_T/\rho_i \ll 1$, where $\rho_i$ is the ion Larmor radius, and
\begin{equation}\label{deltaT}\Delta_T = (\nu_{ei} \omega_T)^{1/2} L_s/(kv_{Te})\end{equation}
is the semi-collisional inner layer thickness~(\ref{deltasc}) with $\omega = \omega_T$. The critical value $\Delta^\prime_{\textrm{crit}} \sim (\hat{\beta}_T/\rho_i)\ln{(\rho_i/\Delta_T)}$. The full definition of $\Delta^\prime_\textrm{crit}$ that is used here to test for diamagnetic stabilization is given in Appendix~\ref{apdx:drifttearing}, which is derived following Ref.~[\onlinecite{connor12}] for electron temperature gradients only~\cite{zocco15}. 

Figure~\ref{fig:growthrates} (middle panel) shows the ratio of $\Delta^\prime$, from Eq.~(\ref{deltaprime}), to $\Delta^\prime_{\textrm{crit}}$, from Eq.~(\ref{deltaprimecrit}), on a logarithmic scale. This ratio is not plotted for $\theta < 20^\circ$, for which $\gamma > \omega_T$ and the strong drift assumption breaks down. The ratio of $\Delta^\prime/\Delta^\prime_\textrm{crit}$ decreases with increasing $\theta$. However, the precise threshold for stabilization ($\Delta^\prime=\Delta^\prime_\textrm{crit}$) only occurs for $\theta=55^{\circ}$ at $t\Omega_{ci} = 70$ and $\theta=57^{\circ}$ at $t\Omega_{ci} = 82$. At $\theta = 35^\circ$, where stabilization is observed, this predicted threshold from boundary layer theory is $13\times$ smaller for $t\Omega_{ci} = 70$ and $30\times$ smaller for $t\Omega_{ci} = 82$.

A similar disagreement between the predictions of boundary layer theory and the measured growth rates of oblique modes has been found previously for the Harris current sheet~\cite{daughton11}. In such an equilibrium, diamagnetic drifts occur only due to density gradients as the temperatures are uniform. Ref.~[\onlinecite{baalrud18}] studied this discrepancy in detail in the collisionless case, concluding that the stabilization is indeed due to electron diamagnetic drift. However, the stabilization was found to be enhanced with respect to the boundary layer theory predictions when the inner tearing layer thickness, $\Delta_\textrm{CLS}$, and the outer ideal region thickness, $\delta$, have insufficient scale separation such that the assumptions of boundary layer theory break down.

Fig.~\ref{fig:growthrates} (bottom panel) shows the ratio of the inner ($\Delta$) to outer region thickness ($\delta$) for the two times on a logarithmic scale. For modes with $\theta < 20^\circ$, we use $\Delta_{SC}$ from Eq.~(\ref{deltasc}) for the inner layer thickness, with $\omega = i\gamma$ from the measured growth rates. For the oblique modes with $\theta \gtrsim 20^\circ$, we take it to be $\Delta_T$ ($\omega = \omega_T$) from Eq.~(\ref{deltaT}). The scale separation between the inner and outer regions is reduced for large oblique angles in a similar manner as seen for the Harris sheet in Fig.~8 of Ref.~[\onlinecite{baalrud18}]. At $\theta=35^{\circ}$, where stability is observed, $\Delta/\delta \approx 0.1$. Although this may seem sufficiently small, similar values in Fig.~8 of Ref.~[\onlinecite{baalrud18}] were large enough to significantly reduce the cut-off angle for oblique modes in the Harris sheet.

The precise reason for the smaller cut-off angle observed here remains an open question. It is conceivable that the combination of electron temperature gradients and breakdown of boundary layer theory could account for this, but further study is required to confirm or reject this explanation. It is significant that two studies~\cite{liu13,akcay16} of collisionless oblique tearing modes in a 1D force-free equilibrium (without temperature gradients) do not find any additional stabilization of oblique modes, as the cut-off angle agrees with the predictions of Eq.~(\ref{collisionlessnormal}). Interestingly, Refs.~[\onlinecite{liu13},\onlinecite{akcay16}] report the growth-rates of oblique modes to be larger than those predicted by boundary layer theory (and even the $\theta=0$ modes) for a range of strong guide fields. 

\subsection{Non-linear phase}

For the linear regime of the plasmoid instability, it is shown in Fig.~\ref{fig:growthrates} (top panel) that the fastest growing modes have small oblique angles ($\theta < 20^\circ$), and that that highly oblique modes with $\theta>35^\circ$ are stabilized. This reduction in the angular distribution of fluctuations may lead one to consider that 2D simulations, which include only the $\theta=0$ modes, may capture the main aspects of this 3D simulation. However, as described in this section, the angular range of fluctuations increases in the non-linear regime.

Figure~\ref{fig:kxthetatime} (bottom panel) shows the angular distribution of fluctuations (at $kd_e=0.133$) also for the non-linear regime of the plasmoid instability, for $88 \lesssim t\Omega_{ci} \leq 140$, which is approximately between the first and second snapshots shown in Fig.~\ref{fig:snapshots}. Over this interval, the trajectories of the white dashed curves in $\theta-t$ from Eq.~(\ref{rotation}) continue to follow the peak values of the fluctuation spectrum, indicating that the mode rotation continues into the non-linear regime while the flux-ropes are not large enough to disrupt the mean properties of the outflow jets. However, beginning at $t\Omega_{ci} \approx 110$, significant power in $P(k_x,k_y,t)$ appears at larger oblique angles than can be expected from mode rotation alone. 

The flux-ropes shown in the second panel of Fig.~\ref{fig:snapshots} show signatures of secondary instabilities. Firstly, on the left side of the domain, at $x\approx L_x/3$, there is evidence of partial coalescence between neighbouring flux-ropes: two flux-ropes visible at the $y=0$ boundary merge into a single flux-rope at $y\approx L_y/2$. Secondly, the flux-ropes show signatures of the kink instability. This is most evident for the flux-rope that forms very close to the flow stagnation point at $x=L_x/2$ and is not monotonically advected downstream by the outflow jets (Fig.~\ref{fig:snapshots} Multimedia view). The safety factor $q(r)=2\pi rB_y/(L_yB_\theta)$ was checked for this flux-rope at $t\Omega_{ci} = 110$ (not shown), soon after it formed, where $r$ is the radial distance from the flux-rope center and $B_\theta(r)$ the poloidal field. For an isolated flux-rope with periodic boundaries, the condition for instability~\cite{kruskal58} requires $q(a)<1$ where $r=a$ is the edge of the flux-rope. It is found that $q(a)=0.6$ at the flux-rope edge (which is taken to be the position of the maximum value of $B_\theta$). Moreover, the kinking of the flux-rope that is visible at $t\Omega_{ci} = 132$ interacts with the reconnection outflow jets and leads to further rotation of the flux-rope as can be seen at times $t\Omega_{ci}=240,320$. This flux-rope grows via reconnection at current sheets on either side to become a ``monster'' flux-rope~\cite{loureirouzdensky16} with diameter as large as $\sim L_z/3$ ($18d_{i0}$) by the end of the simulation at $t\Omega_{ci}=400$. 
 
It should be noted that although secondary flux-ropes are observed at late times (e.g. $t\Omega_{ci}=320$), there are relatively few compared with those forming along thin separatrix current layers in the 3D collisionless simulations of Ref.~[\onlinecite{daughton11}]. The separatrix current layers in the present simulation appear less intense than those in Ref.~[\onlinecite{daughton11}], presumably due to collisional broadening.

\subsection{\label{sec:heattransport}Stochastic magnetic field and heat transport}

\begin{figure*}
\includegraphics[width=0.75\textwidth]{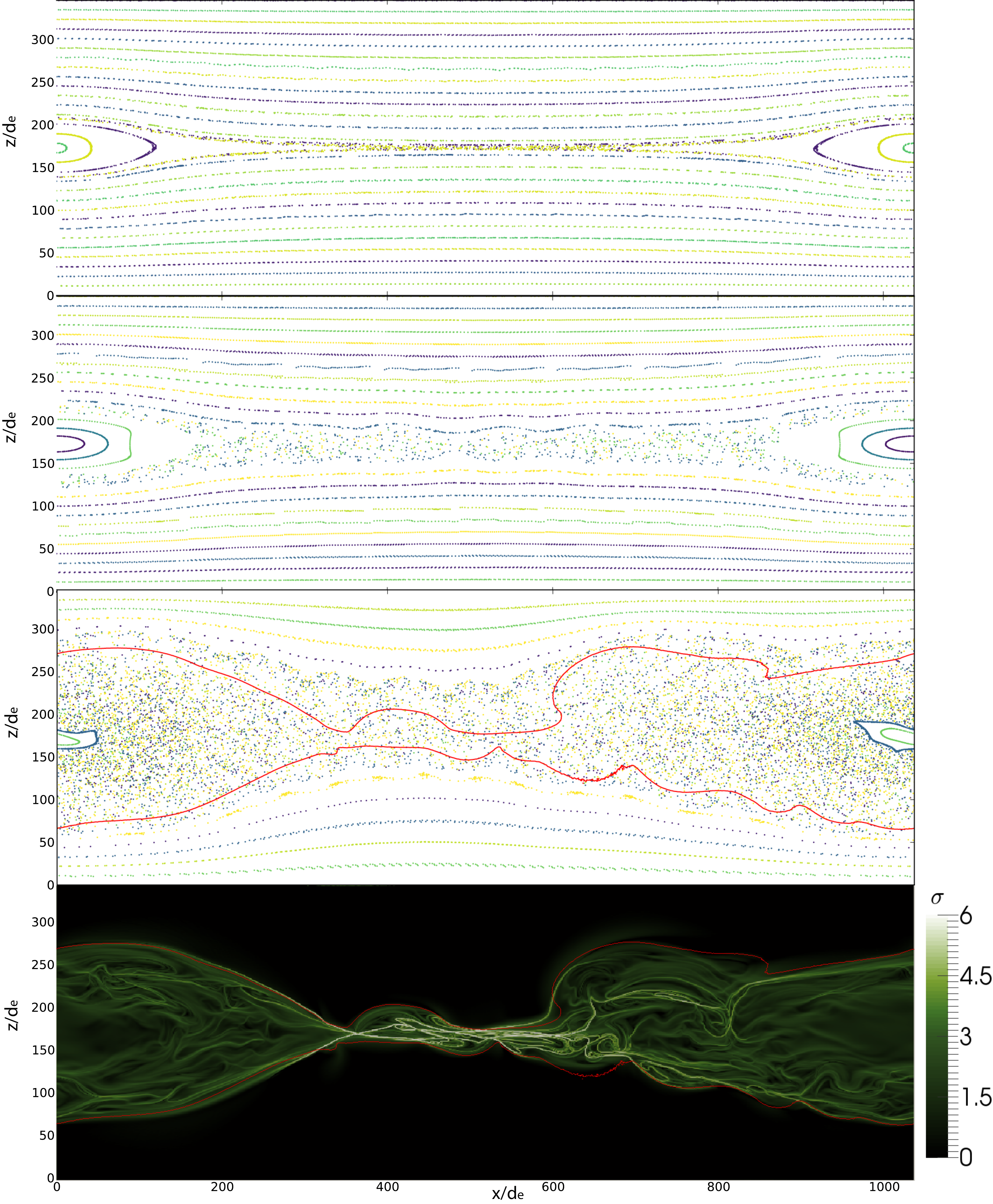}
\caption{\label{fig:poincare}Poincar\'e plots showing the intersection of magnetic field-lines with the $y=L_y/2$ surface at $t\Omega_{ci}=88$ (first panel), $t\Omega_{ci}=132$ (second panel) and $t\Omega_{ci}=396$ (third panel). Field-lines are traced $1000$ times through the simulation domain (based on $L_x$) using a volume preserving integration scheme~\cite{finnchacon05}. The fourth panel shows the exponentiation factor $\sigma$ at $t\Omega_{ci}=396$, calculated by tracing field-lines a distance $L_s=L_y/2$ from a plane of seed points at $y=L_y/2$. Also shown in the third and fourth panels is a red contour of the electron temperature with $T_e/T_{e0} = 1.15$. }
\end{figure*}

In 3D, the formation of oblique plasmoids at multiple resonant surfaces can lead to the breakdown of magnetic surfaces. Since plasma transport is primarily along the magnetic field, the mixing of magnetic field-lines can lead to enhanced plasma mixing. In light of the above discussion on the stabilization of strongly oblique modes in the linear regime, and on the secondary instabilities in the non-linear regime, it is useful to briefly characterize the extent of any stochastic magnetic field regions and their role in plasma transport. 

Figure~\ref{fig:poincare} (top three panels) shows Poincar\'e surfaces of section with magnetic field-lines for different times during the simulation. Here, the field-lines are integrated a distance of $1000 L_x$ through the simulation domain, crossing through the periodic boundaries in the $x$ and $y$-directions, and the surface of section is the plane at $y=L_y/2$. To reliably integrate the field-lines over such a distance, we use a volume preserving method~\cite{finnchacon05} that ensures $\boldsymbol{\nabla}\cdot \boldsymbol{B} = 0$ to numerical round-off and has been shown to well reproduce boundaries between domains of ordered and stochastic magnetic field~\cite{ciaccio13}.

Figure~\ref{fig:poincare} (top) shows a Poincar\'e section at $t\Omega_{ci} = 88$, which is at the start of the non-linear phase of the oblique plasmoid instability. As well as the upstream unreconnected flux, there are clearly visible magnetic flux-surfaces in the downstream region showing a magnetic island. This island is seeded in the single mode perturbation of Eq.~(\ref{initialpert}) and remains stable as it grows due to the quasi-2D nature of the Sweet-Parker reconnection. In between the upstream and downstream flux-surfaces is a thin region of stochastic magnetic field caused by the overlap of oblique magnetic flux-ropes. At the later times of $t\Omega_{ci} = 132$ (second panel) and $t\Omega_{ci}=396$ (third panel) the size of the stochastic region increases until it fills a significant part of the simulation volume at saturation. Within this middle volume there is no indication of any structure, suggesting that the ``flux-ropes'' that are visible in Fig.~\ref{fig:snapshots} do not confine magnetic field-lines over such large distances. 

To compare the regions of magnetic field mixing with plasma mixing, we consider the electron temperature $T_e$. The red contour in Fig.~\ref{fig:poincare} is $T_e/T_{e0} = 1.15$, just above the background value. Although this contour covers a significant part of the stochastic region, there are clear regions where the magnetic field is stochastic outside of this contour (choosing lower threshold values for the contour do not give better agreement). This result is similar to the test-particle study of Ref.~[\onlinecite{borgogno17}], where the electron mixing region was found to be somewhat smaller than the stochastic magnetic field region. 

In the present simulation, where the plasma and fields are self-consistently coupled, the finite electron velocity may limit the spread of electrons along the full volume of the stochastic region. To test this, we plot the magnetic field line exponentiation factor $\sigma$, which measures the exponential rate of separation of neighbouring magnetic field-lines~\cite{boozer12,daughton14,le18}. It is defined as 
\begin{equation}\sigma = \ln(\rho_{\textrm{max}}^{1/2}),\end{equation}
where $\rho_\textrm{max}$ is the maximum eigenvalue of the Cauchy-Green deformation tensor $(\nabla_{\boldsymbol{x}_0}\boldsymbol{x}_f)(\nabla_{\boldsymbol{x_0}}\boldsymbol{x}_f)^T$ of the field-line mapping $\boldsymbol{x}_0\rightarrow \boldsymbol{x}_f(\boldsymbol{x}_0)$. Here $\boldsymbol{x}_0$ are taken to be an array of seed points in the $y=L_y/2$ surface and $\boldsymbol{x}_f$ are the final positions after integrating a distance $L_s=L_y/2$ along the magnetic field-lines. The exponentiation factor is similar to the squashing degree $Q$, used to define quasi-separatrix layers~\cite{titov02,priest95}, and the finite time Lyapunov exponent often used to characterize fluid flows.

Fig.~\ref{fig:poincare} shows that the region of significant $\sigma$ agrees more closely with the electron temperature contour than the stochastic region shown in the Poincar\'e plot. The maximum $\sigma = 6.7$ at this time. We also find that the agreement between the $T_e$ contour and the region of significant $\sigma$ is fairly close for most of the simulation (not shown), apart from at early time where there is rapid change in $T_e$ due to Joule heating in the current layer. This suggests that the snapshots of $T_e$ in Fig.~\ref{fig:snapshots} trace out the magnetic topology to a reasonable degree, but due to the finite electron velocity they do not explore the whole stochastic region instantly.

\section{\label{sec:kinetictransition}Transition to kinetic reconnection}

\begin{figure*}
\includegraphics[width=0.8\textwidth]{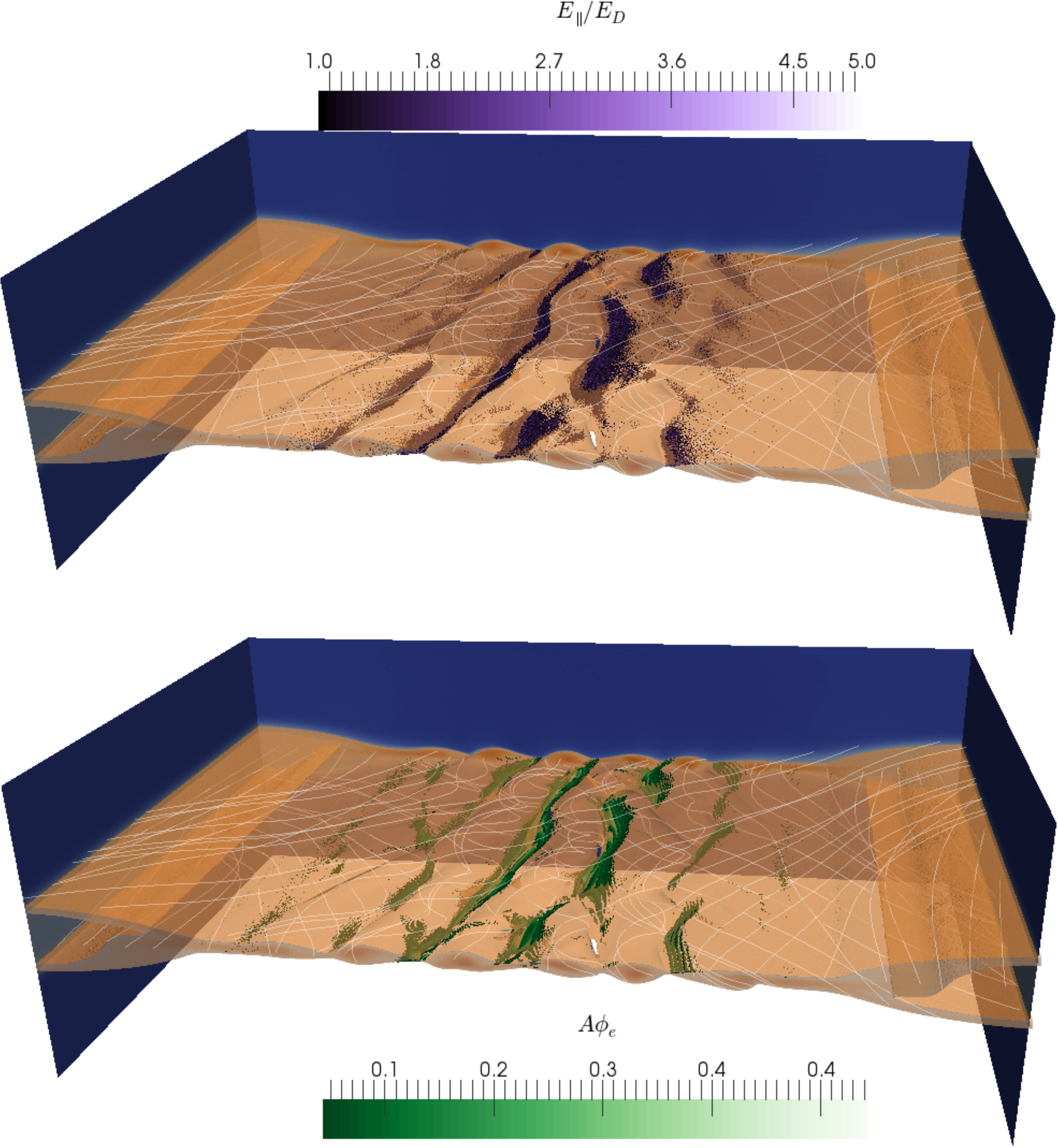}
\caption{\label{fig:epar-agyro} Top panel: Ratio of the local parallel electric field to the Dreicer field, shown for $E_\parallel/E_D \geq 1$ (purple). Bottom: Agyrotropy~\cite{scudder08}, a scalar measure of the departure of the electron pressure tensor from cylindrical symmetry, shown for $A\phi_e \geq 0.08$. Both panels show a contour of the electron temperature ($T_e=0.08$) and magnetic field-lines (white) to depict the oblique flux-ropes.  }
\end{figure*}

The transition from collisional to kinetic reconnection has been previously studied using 2D first-principles simulations in Refs.~[\onlinecite{daughton09a,daughton09b,roytershteyn10}]. When the current layer thickness falls below the ion kinetic scale (either by thinning of laminar layers, or by new layers forming between magnetic islands due to the plasmoid instability), the reconnection electric field is observed to become larger than the critical Dreicer threshold $E_D$. This triggers rapid thinning of the current layer until it reaches electron kinetic scales ($\approx 2 d_e$)~\cite{daughton09b}. As this occurs, resistive friction is no longer sufficient to balance the electric field and it is instead supported by gradients in the off-diagonal elements in the electron pressure tensor at the X-point location~\cite{roytershteyn10}. The previous studies report these results for measurements taken at a single point in space - the primary X-point of the 2D reconnection layer. In this section, the physics of this transition is examined for the current simulation, with focus on the 3D spatial locations where signatures of kinetic reconnection occur.  
 
Figure~\ref{fig:epar-agyro} (top) shows an isovolume of $E_{\parallel}/E_D \geq 1$ (purple), the ratio of the parallel electric field to the critical Dreicer field at $t\Omega_{ci}=132$. The magnetic surfaces are indicated by a contour of $T_e$ (orange), which shows the flux-ropes have grown large enough to break-up the primary current layer. Intense current-layers that form between the flux-ropes are found to reach thicknesses on the electron kinetic scale $\delta \approx 1.5 d_e$ (not shown), in agreement with the findings of the previous 2D studies. The spatial locations of the super-Dreicer parallel electric fields are in good agreement with the locations of these thin current layers, and reach values as large as $E_\parallel/E_D = 5$.

The bottom panel of Fig.~\ref{fig:epar-agyro} shows an isovolume of the electron pressure agyrotropy with $A\phi_e\geq 0.08$ (green). This agyrotropy is a scalar measure of the departure of the pressure tensor from cylindrical symmetry about the magnetic field~\cite{scudder08}, and significant values of $A\phi_e$ have been observed in both simulations and spacecraft data~\cite{scudder12} at sites of collisionless magnetic reconnection. The isovolume of $A\phi_e \geq 0.08$ also appears to be spatially co-located with the isovolume of $E_\parallel/E_D\geq 1$, and the intense current layers that form between the flux-ropes. More quantitatively, there is a moderate positive correlation between $E_\parallel/E_D$ and $A\phi_e$ (Pearson coefficient $0.6$) in regions where the electric field is super-Dreicer, $E_\parallel/E_D\geq 1$. The primary mechanism for the generation of the agyrotropic electron distributions is presently unclear, although several possibilities have been suggested based upon tracking particles in simulations of collisionless reconnection with strong electric field gradients~\cite{wendel16}. Since electron collisions act to isotropize the pressure tensor, significant $A\phi_e$ is taken here to be a signature of the transition to kinetic reconnection. 

\begin{figure}
\includegraphics[width=0.5\textwidth]{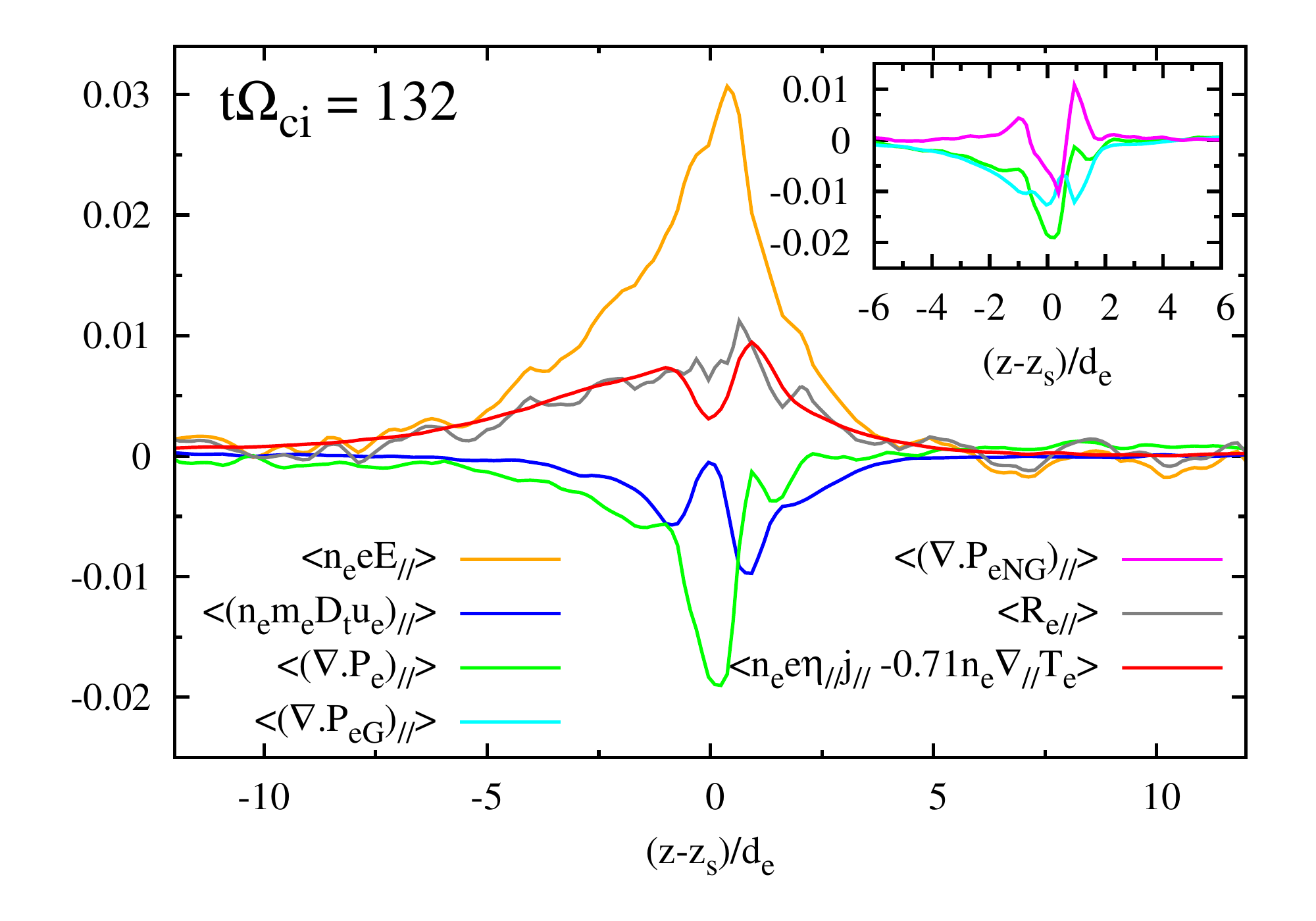}
\caption{\label{fig:ohmslawsafter}Contributions to parallel force balance at $t\Omega_{ci}=132$. Contributions have been averaged in time and along the field-lines - see text for definitions.  Quantities are expressed in ion units, after normalization by $n_0B_{r0}v_{A}$. Inset: Decomposition of electron pressure tensor (green) into gyrotropic (cyan) and non-gyrotropic (magenta) components.}
\end{figure}

Figure~\ref{fig:ohmslawsafter} shows the electron momentum balance at $t\Omega_{ci0} = 88$, from a line of seed points at $\boldsymbol{x}_0 = (428,0,z)$ for $z\in [z_s-12,z_s+12]$, $z_s=L_z/2$, and integrated a distance of $L_s=10$ ($d_e$). This small value of $L_s$ was chosen to prevent apparent broadening of the non-ideal electric field region when integrating along stochastic field-lines that exit the kinetic scale diffusion regions, but still have $E_\parallel \neq 0$ due to finite plasma resistivity. In contrast to Fig.~\ref{fig:ohmslaws}, the region with finite $<n_e e E_\parallel>$ (orange) is significantly thinner, with half-thickness $\approx 1.5$ ($d_e$). The electron pressure tensor term (green) is now the largest balancing the electric field term at the center of the current layer. To examine this further, inset shows the break-down of the electron pressure term into the gyrotropic (cyan) and non-gyrotropic (magenta) components, where $\boldsymbol{P}_{eG} = P_{e\parallel}\boldsymbol{\hat{b}}\boldsymbol{\hat{b}} + P_{e\perp}\left(\mathbb{I}-\boldsymbol{\hat{b}}\boldsymbol{\hat{b}}\right)$, $\boldsymbol{P}_{eNG} =  \boldsymbol{P}_{e}-\boldsymbol{P}_{eG}$, $P_{e\parallel} = \boldsymbol{P}_{e}:\boldsymbol{\hat{b}}\boldsymbol{\hat{b}}$ and $P_{e\perp} = \left(\textrm{Tr}[\boldsymbol{P}_{e}]-P_{e\parallel}\right)/2$. The non-gyrotropic part is significant in a thin region $\delta \approx \rho_e < d_e$ in the very center of the current layer, but on either side the gyrotropic part has a larger contribution to $<n_e e E_\parallel>$. This gyrotropic part has been observed in 3D collisionless reconnection simulations in Refs.~[\onlinecite{liu13}],~[\onlinecite{sauppe18}], although it is identically zero at the X-point in 2D simulations due to symmetry. As such, the role of this term in decoupling electrons from magnetic field-lines and permitting reconnection remains unclear. Strictly, $E_\parallel \neq 0$ is not a sufficient condition for reconnection in 3D, where the condition $\boldsymbol{\nabla}\times (\boldsymbol{E}+\boldsymbol{u}_e\times \boldsymbol{B})\neq 0$ is more appropriate~\cite{hornig07}. Only the part of the gyrotropic term with non-zero curl is able break the electron frozen-in condition and determine the electron diffusion region but, unfortunately, the noise level in the components of the non-ideal electric field is too large to reliably compute the derivatives needed to examine this issue. Nevertheless, the $\approx 1.5 d_e$ half-thickness of the non-ideal region remains in good agreement with 2D collisionless simulations, as well as the significant role of both the electron inertia and non-gyrotropic pressure tensor terms in balancing the non-ideal electric field. These results are taken as confirmation of the transition to kinetic reconnection, which occurs within these thin current layers on the electron kinetic scale.

\section{\label{sec:conclude}Summary and discussion}

The transition from collisional to kinetic magnetic reconnection was studied for the first time in 3D, using a first-principles kinetic approach with a Monte-Carlo treatment of the Fokker-Planck collision operator. Initial reconnection in the low-$\beta$ force free current sheet proceeded in a quasi-2D Sweet-Parker regime, keeping the symmetry of the initial perturbation, as expected for the ``single X-line collisional'' region of the reconnection phase diagram~\cite{jidaughton11} shown in Fig.~\ref{fig:phasediagram}.

In the low-$\beta$ sheet, intense Joule heating leads to more rapid thinning than reported for previous studies~\cite{daughton09a,daughton09b} with $\beta \approx 1$. While the current layer remains collisional, transport resulting from the kinetic description of collisions can include the classical effects of temperature dependent and anisotropic resistive and thermal friction, viscosity, heat conduction, and species thermal equilibration. However, a simplified resistive-MHD model that includes a Spitzer-type law for the resistivity, neglects heat conduction, and assumes equal ion and electron temperatures was found to reasonably well reproduce the current layer thinning profile for this simulation. Prior to disruption of the current layer, the resistive thinning causes the simulation to transition to the ``Multiple X-line'' hybrid region of Fig.~\ref{fig:phasediagram}. 

The 2D symmetry of the initial phase was broken by the oblique plasmoid instability, which occurred in the dynamically thinning Sweet-Parker sheet with well established reconnection outflow jets. In the early phase of the plasmoid instability, the tearing modes were found to be in the semi-collisional regime (with growth rates smaller than the collision frequency, $\nu_{ei}>\gamma$, and an inner layer thickness below the sound-radius, $\Delta_{SC} < \rho_s$). The growth rates for modes with small oblique angles ($\theta \lesssim 20^{\circ}$), which agreed well with linear semi-collisional theory predictions~\cite{drakelee77}, were found to be large compared to the rates of mode stretching and rotation by the outflow jets.

However, strongly oblique modes were stabilized at a much lower angular cut-off ($\theta \approx 35^{\circ}$) than predicted for standard tearing modes. The presence of electron temperature gradients from the Joule heating was considered as a mechanism for this observed stabilization, but a theory accounting for this physics~\cite{connor12} was also found to underpredict the amount of stabilization. The precise reason for the stabilization observed in the present simulation remains an open question, and it is possible that the validity of boundary layer theory is violated for the strongly oblique modes~\cite{baalrud18}. 

Despite this narrow angular spectrum of oblique modes in the linear regime, magnetic energy is subsequently injected into oblique fluctuations by a combination of flux-rope rotation by the reconnection outflow jets, and secondary kink and coalescence instabilities. A region of stochastic magnetic field is formed by the plasmoid instability, which grows over time as more flux is reconnected, and agrees reasonably well with the observed extent of electron heat transport. 

Apart from long wavelength variations in the $y$-direction, the transition to kinetic reconnection proceeds in a manner similar to 2D simulations~\cite{daughton09a,daughton09b}. The parallel electric field becomes super-Dreicer ($1 \lesssim E_\parallel/E_D \lesssim 5$) at kinetic-scale current layers that form between the oblique flux-ropes, and a significant part of the parallel force is balanced by electron pressure tensor and inertia terms (although the former has a gyrotropic component~\cite{liu13} not present in 2D). Secondary flux-ropes are observed to form in these thin current layers at late time in the simulation. The overall behavior described for this 3D simulation supports the picture of the plasmoid mediated transition to kinetic reconnection in the ``Multiple X-line'' hybrid regime, as indicated in Fig.~\ref{fig:phasediagram}.  

Solar flare reconnection, which occurs in low-$\beta$ force free current layers, is also argued to be in the ``Multiple X-line hybrid'' regime based on present understanding. With a flare Lundquist number of $S\sim 10^{13}$ and system-size~\cite{jidaughton11} $\lambda = L/\delta_i \sim 4\times 10^7$, direct numerical simulation is unfeasible in the near future and we are left to extrapolate from smaller simulation and experimental studies. The $S\sim 10^{3-4}$ and $\lambda \sim 10^{2-3}$, as well as the low-$\beta$ initial conditions used for this paper are relevant to the newly constructed Facility for Laboratory Reconnection Experiments (FLARE~\cite{jiflare18}). 

Recently, a number of laboratory magnetic reconnection experiments have observed the break up of current layers due to the plasmoid instability~\cite{olsen16,jaraalmonte16,hare17}, but the plasmoid mediated transition from collisional to kinetic reconnection has not yet been studied in detail. The full picture of the plasmoid instability in FLARE should take into account the resistive thinning of the Sweet-Parker layer that forms due to inductive current drive, the influence of the flux-core boundary conditions on the growth of oblique modes, the semi-collisional inner layer physics, and the role of outflow jets in the stretching and rotation of modes. It may also require the consideration of ion-neutral and neutral-neutral collisions~\cite{jaraalmonte19}. Future work will extend the present study to experimentally realistic cylindrical geometry of the FLARE experiment, including the relevant physics, to better enable comparisons to be drawn.

\begin{acknowledgments}
This work is supported by the Basic Plasma Science Program from the U.S. Department of Energy, Office of Fusion Energy Sciences. The large simulation was performed at the National Energy Research Scientific Computing Center (NERSC), a U.S. Department of Energy Office of Science User Facility operated under Contract No. DE-AC02-05CH11231. Supporting simulations used resources from the Los Alamos National Laboratory Institutional Computing Program, which is supported by the U.S. Department of Energy National Nuclear Security Administration under Contract No. DE-AC52-06NA25396.
\end{acknowledgments}

\appendix

\section{\label{apdx:drifttearing}$\Delta_{\textrm{crit}}^\prime$ due to temperature gradients}

Ref.~[\onlinecite{connor12}] gives a theory for the semi-collisional drift-tearing and internal kink instabilities for arbitrary plasma-$\beta$ and $\Delta^{\prime}$, including ion-orbit effects via a gyrokinetic treatment. In general the dispersion relations need to be computed numerically, but closed form expressions can be found in certain asymptotic limits. Firstly, the semi-collisional theory of Eq.~(\ref{semicollisionalnormal}) can be found~\cite{zocco15} in the limit of cold ions and small-$\Delta^\prime$. Secondly, a dispersion relation can be found for the strong drift regime (including finite ion orbits) by expanding $\omega = \omega_r + i\gamma$ in powers of $(\Delta/\rho_i)\ll 1$, where $\Delta$ is the semi-collisional layer thickness and $\rho_i$ is the ion gyroradius. Including only electron temperature gradients~\cite{zocco15} so $\Delta=\Delta_T$ defined in Eq.~(\ref{deltaT}), and neglecting density and ion temperature gradients, the lowest order frequency $\omega_0$ is real:
\begin{equation}\frac{\omega_0}{\omega_T} = \frac{1.71 \sqrt{1+\tau}}{\sqrt{1+\tau} + \sqrt{2.13 \tau}},\end{equation}
where $\tau = T_e/T_i$, and $\omega_T = k T_e/(eBL_T)$. At the next order the growth rate scales as 
\begin{equation}\frac{\gamma}{\omega_T} \sim \frac{\Delta_T}{\pi \hat{\beta}_T} \left[\Delta^\prime - \Delta^\prime_{\textrm{crit}}\right],\end{equation}
where $\beta_T = (\beta_e/2) L_s^2/L_T^2$. The critical threshold for instability, $\Delta^\prime_{\textrm{crit}}$, is given by the expression
\begin{equation}\label{deltaprimecrit}\Delta^\prime_{\textrm{crit}} = \frac{\sqrt{\pi}\hat{\beta}_T}{\rho_i} \frac{\omega_0^2}{\omega_T^2} \frac{\tau}{(1+\tau)^2}\ln\left[\frac{\rho_i}{\Delta_T} \sqrt{\frac{\omega_T}{2\omega_0}}\right] - \frac{\pi \hat{\beta}_T}{\rho_i} \frac{\omega_0^2}{\omega_T^2} \bar{I}(\tau).\end{equation}
Here, the integral $\bar{I}(\tau)$ is from the gyrokinetic ions~\cite{connor12,zocco15}. It is given by
\begin{equation}\bar{I}(\tau) = \int_0^{\infty} dk \left[\frac{F(k)}{G(k)} - \frac{\tau}{1+\tau} + \frac{\tau}{\sqrt{\pi} (1+\tau)^2 (1+k)}\right], \end{equation}
with
\begin{equation}F(k) = \tau \left[\exp{(-k^2/2)}I_0(k^2/2)-1\right],\end{equation}
$I_0$ is the modified Bessel function of the first kind, and $G(k) = F(k)-1$. To calculate $\Delta^\prime_{\textrm{crit}}$ that is used in Fig.~\ref{fig:growthrates}, this integral is calculated numerically based upon the local $\tau$ at each rational surface. The integral is negative (it is stabilizing), and has typical value $\bar{I} \approx -0.5$ for the parameters used here.


%

\end{document}